\documentclass{article}
\usepackage[utf8]{inputenc}
\usepackage{authblk}
\usepackage{setspace}
\usepackage[top = 1.25in, bottom = 1.25in, left=1in, right = 1in]{geometry}
\usepackage{graphicx}
\graphicspath{ {./figures/} }
\usepackage{subcaption}
\usepackage{array}
\usepackage[
citestyle=numeric-comp,
sorting = none,
url=false,
autocite = superscript
]{biblatex}
\AtEveryBibitem{%
  \clearfield{note}%
}
\newcommand{\citeSupp}[1]{\textsuperscript{[ref.}\autocite{#1}\textsuperscript{]}}
\usepackage{comment}
\usepackage{multicol}
\usepackage{enumitem}
\usepackage{supertabular}
\usepackage{amsmath}
\usepackage{lineno}
\usepackage{graphicx}%
\usepackage{multirow}%

\usepackage{amsmath,amssymb,amsfonts}%
\usepackage{amsthm}%
\usepackage{makecell}
\usepackage{mathrsfs}%
\usepackage[title]{appendix}%
\usepackage{xcolor}%
\usepackage{textcomp}%
\usepackage{tabularray}
\usepackage{manyfoot}%
\usepackage{booktabs}%
\usepackage{float}
\restylefloat{table}
\usepackage[parfill]{parskip}
\usepackage{algorithm}%
\usepackage{algorithmicx}%
\usepackage{algpseudocode}%
\usepackage{listings}%

\title{A Fast Analytical Model for Predicting Battery Performance Under Mixed Kinetic Control}

\author[1]{Hongxuan Wang}
\author[1]{Fan Wang}
\author[1*]{Ming Tang}

\affil[1]{Department of Materials Science and NanoEngineering, Rice University, Houston, 77005, TX, USA}
\affil[*]{Correspondence: mingtang@rice.edu}

\begin{document}

\maketitle

\begin{abstract}
The prediction of battery rate performance traditionally relies on computation-intensive numerical simulations. While simplified analytical models have been developed to accelerate the calculation, they usually assume battery performance to be controlled by a single rate-limiting process, such as solid diffusion or electrolyte transport. Here, we propose an improved analytical model that could be applied to battery discharging under mixed control of mass transport in both solid and electrolyte phases. Compared to previous single-particle models extended to incorporate the electrolyte kinetics, our model is able to predict the effect of salt depletion on diminishing the discharge capacity, a phenomenon that becomes important in thick electrodes and/or at high rates. The model demonstrates good agreement with the full-order simulation over a wide range of cell parameters and offers a speedup of over 600 times at the same time. Furthermore, it could be combined with gradient-based optimization algorithms to very efficiently search for the optimal battery cell configurations while numerical simulation fails at the task due to its inability to accurately evaluate the derivatives of the objective function. The high efficiency and the analytical nature of the model render it a powerful tool for battery cell design and optimization.  
\end{abstract}


\section{Introduction}
The popularity of rechargeable lithium-ion batteries (LIB) has experienced an explosive growth since their first commercialization in the 1990s. Compared to the other energy storage systems, LIBs exhibit relatively high energy density and decent cycle life, enabling a wide range of applications from consumer electronics to electric mobility. With the demand for LIBs increasing rapidly because of the wider adoption of electrical vehicles and grid energy storage systems, the development of next-generation LIBs with significantly improved energy density, rate performance, safety, cycling stability and fast-charging capability is highly sought-after. The success of this effort requires continued optimization of battery structure across different length scales from the particle to the cell and pack levels. For instance, the replacement of polycrystalline LiNi$_{1-x-y}$Mn$_x$Co$_y$O$_2$ with single crystalline particles could lead to significantly better cycle life and higher tap density\autocite{qian_single-crystal_2020, wang_single_2020, huang_pulse_2023}, and the design and fabrication of thick battery electrodes, some with three-dimensional architecture\autocite{li_fabrication_2019, sun_hierarchical_2019,chu_3d_2021, zhang_multiscale_2021, zhang_tunable_2021} to facilitate ionic transport, are under intense study because they could increase the energy density of batteries by increasing the amount of active materials relative to the inactive components\autocite{kuang_thick_2019, gallagher_optimizing_2015, zheng_comprehensive_2012}. 

The large design parameter space inherent in LIBs makes battery modeling an essential tool for their design and optimization to unravel the complex dynamics and reduce the testing turnaround time. 
While battery simulations that explicitly resolve the electrode microstructure have gained traction in recent years and could provide valuable insights \autocite{garcia_microstructural_2005, danner_thick_2016, malik_complex_2022}, their high computation cost still prevents their extensive use.
Instead, a standard modeling approach is numerical simulations based on the porous electrode theory pioneered by Newman and coworkers\autocite{doyle_modeling_1993, fuller_simulation_1994, thomas_mathematical_2002, ferguson_nonequilibrium_2012, newman_electrochemical_2021}. These simulations are commonly referred to as pseudo-two-dimensional (P2D) simulations as they couple one-dimensional electrolyte transport at the macroscopic electrode level and one-dimensional lithium diffusion within the solid phase at the microscopic particle level. P2D simulations provide a comprehensive description of the reaction-transport kinetics in battery cells, but the highly-coupled governing equations are computationally expensive to solve. While various numerical techniques have been employed to accelerate P2D simulations, such as proper orthogonal decomposition\autocite{cai_reduction_2008} and orthogonal collocation\autocite{cai_lithium_2012, bizeray_lithium-ion_2015}, it remains challenging to use this approach to explore the large parameter space or perform large-scale optimization. Another common approach involves building equivalent circuit models (ECMs) for LIBs\autocite{yann_liaw_modeling_2004, buller_impedance-based_2005, hu_comparative_2012}. This approach requires empirically fitting the parameters in ECMs to experimental data to reflect the inner workings of the battery cell. Nevertheless, the lack of physical insight makes such models prone to errors when extrapolated to conditions beyond the scope of the fitting data. Because the model parameters do not translate easily to materials properties, it is also difficult to use ECMs to optimize the battery structure. As an intermediate between the P2D simulation and the ECMs, simplified physics-based models are another candidate for battery modeling. These models are more efficient to solve than the full-order numerical simulations and also provide direct insight into the structure-property relation of batteries than the (semi-)empirical models.

In simplified physics-based models, the system complexity is typically  reduced by assuming a predominant rate-limiting process. For example, the widely used single particle model (SPM) assumes that the electrolyte transport is facile and all of the active material particles are uniformly (de)lithiated. As such, the (dis)charging behavior of the electrode is approximately simulated by solving the solid diffusion equation for a single particle coupled with the charge transfer process at the particle/electrolyte interface\autocite{doyle_analysis_1997, liu_analytical_2006, han_simplification_2015}.  
On the other end, Wang and Tang recently proposed an analytical model to predict the battery rate performance when it is controlled by the electrolyte transport\autocite{wang_quantitative_2020}. In this kinetic regime, the reduction of usable capacity with increasing discharging rates is caused by the lithium salt depletion in the electrolyte. As a result, active material particles in the salt-depletion region could not be fully lithiated, leading to incomplete discharge. 
The Wang-Tang model assumes fast solid diffusion and that the active material particles are either completely unlithiated or lithiated depending on whether they reside outside the salt depletion zone (DZ) or not. 
Reduction of the model complexity is realized by the assumptions of the steady-state electrolyte transport and characteristic reaction distributions. 
The discharge capacity is estimated from the width of the electrolyte penetration zone $L_\textrm{PZ}$, in which the salt concentration is non-zero, as $L_\textrm{PZ}/L_{cat}$, where $L_{cat}$ is the cathode thickness. 

While the aforementioned models focusing on a single rate-limiting step have proven effective in their respective applicable regimes, the (dis)charge behavior of LIBs is often determined jointly by the solid diffusion and electrolyte transport kinetics, which renders such models insufficient. 
The accuracy of SPM becomes unsatisfactory when the electrodes are thick and/or the discharge rate is large, where electrolyte diffusion becomes relatively sluggish.  
The Wang-Tang model is prone to larger errors when the active material particle size is large and the solid diffusion process could no longer be neglected.
With the relentless pursuit for better battery performance and economics, these scenarios are becoming increasingly relevant as the industry is pushing for thicker electrodes to enhance the energy density and the manufacturing efficiency also favors the use of larger particles to increase the tap density.

A number of models have been developed in recent years to extend the SPM to consider electrolyte dynamics\autocite{khaleghi_rahimian_extension_2013, luo_new_2013, moura_battery_2017, perez_optimal_2017, marquis_asymptotic_2019}, which is usually achieved by assuming certain simplified forms of the reaction distribution within the electrode.
Moura et al. derived an extended SPM that incorporates the electrolyte phase potential drop into the cell voltage based on the assumption of uniform Li intercalation flux\autocite{moura_battery_2017}. In another extension, Rahimian et al. approximate the electrolyte concentration and potential with third-order polynomials and fit the polynomial coefficients to full-order P2D simulations\autocite{khaleghi_rahimian_extension_2013}. Luo et al. extended the SPM in a different approach, which assumes that the spatial distribution of the open circuit potential (OCP) of active material particles could be described by an exponential function\autocite{luo_new_2013}. The parameters in the function are calculated from fitting to the OCPs of three representative particles at different thickness of the electrode.  
While these extended SPM demonstrate improved predictions, they could not be applied to the situation where salt depletion develops in the electrolyte, which becomes significant at large electrode thickness and/or (dis)charging rates. 
The need for parameter fitting in some of the models also make them less general and solvable by numerical methods only.  

In this work, we propose an improved physics-based analytical model for predicting the battery discharge process under the mixed kinetic control of salt depletion and solid diffusion. The model is suitable for active materials that display solid solution behavior upon (dis)charging, such as the layered transition metal oxides Li(Ni$_{1-x-y}$Mn$_{x}$Co$_y$)O$_2$ (NMC) or Li(Ni$_{1-x-y}$Co$_{x}$Al$_y$)O$_2$ (NCA) that are widely used in commercial LIBs. 
Because the OCP of these materials is sensitive to the lithium stoichiometry or state of charge (SoC), the lithium intercalation flux tends to be more homogeneously  distributed across the electrode than the phase-changing electrode materials such as LiFeO$_4$ (LFP) and Li$_4$Ti$_5$O$_{12}$ (LTO),
which exhibit a propagating reaction front. 
Therefore, they are idealized as the \textbf{uniform-reaction}-type (UR) electrodes in the Wang-Tang model. 
The current model consists of an electrolyte transport and a solid phase module to capture the two coupled processes.
The electrolyte module extends the Wang-Tang model to handle concentration-dependent electrolyte properties and spatially-varying electrode properties (e.g. porosity). 
It is used to predict the width of the electrolyte penetration zone (PZ) and the spatial distributions of the salt concentration $c_l$ and electrolyte potential $\Phi_l$ within the PZ. 
The new solid phase module calculates the 
OCP distribution of the active material in the PZ, and estimates the average lithium concentration $c_s$ in electrode particles via the solution to the solid diffusion equation.
The cell-level depth of discharge (DoD) is then calculated as a function of the cell voltage from the integration of $c_s$ in space.  
Compared to the original UR model\autocite{wang_quantitative_2020}, the new model not only addresses the effect of solid diffusion on the rate performance but also predicts the discharge voltage curve in the presence of salt depletion.
We name it the URCs model, recognizing its applicability to UR-type electrodes and the additional calculation of the lithium concentration distribution in the solid phase. 

The URCs model is compared with P2D simulations over a wide range of cell parameters to examine its performance. The discharge capacity and energy output predicted by the model exhibit very good agreement with the simulated results with an average error less than 10\%, excluding the low DoD regime in cells containing the graphite anode, which is known for its non-UR behavior. 
At the same time, it offers a speedup of over 600 folds versus the state-of-the-art P2D solvers. More significantly, the URCs model, unlike numerical simulation, is able to work in synergy with gradient-based optimization algorithms to efficiently locate optimal battery cell parameters because it permits the accurate evaluation of the objective function gradients. A hybrid global optimization scheme, which employs the URCs model for rapid scan of the parameter space and P2D simulation for refined local search, is demonstrated with both speed and accuracy. These advantages render the URCs model a useful tool for battery structure design as well as onboard applications.

\section{Results}
\subsection{Uniform-Reaction (UR) and Moving-Zone-Reaction (MZR) Behavior}\label{urtype}
In this section, we provide a concise review of the two distinct types of reaction distribution within electrodes that underpin the simplifying assumptions in the Wang-Tang model~\autocite{wang_quantitative_2020}: the uniform-reaction (UR) and the moving-zone-reaction (MZR) behavior. UR and MZR behaviors manifest in battery chemistries with OCPs that are  sensitive and insensitive to lithium stoichiometry, respectively. Cathode materials such as solid-solution-like transition metal oxides, including NMC and NCA, commonly exhibit UR behavior, while materials that undergo pronounced first-order phase transitions  during (de)lithiation, exemplified by LFP, typically demonstrate MZR behavior.

\begin{figure}[!tbh]
    \centering
    \includegraphics[width = \textwidth]{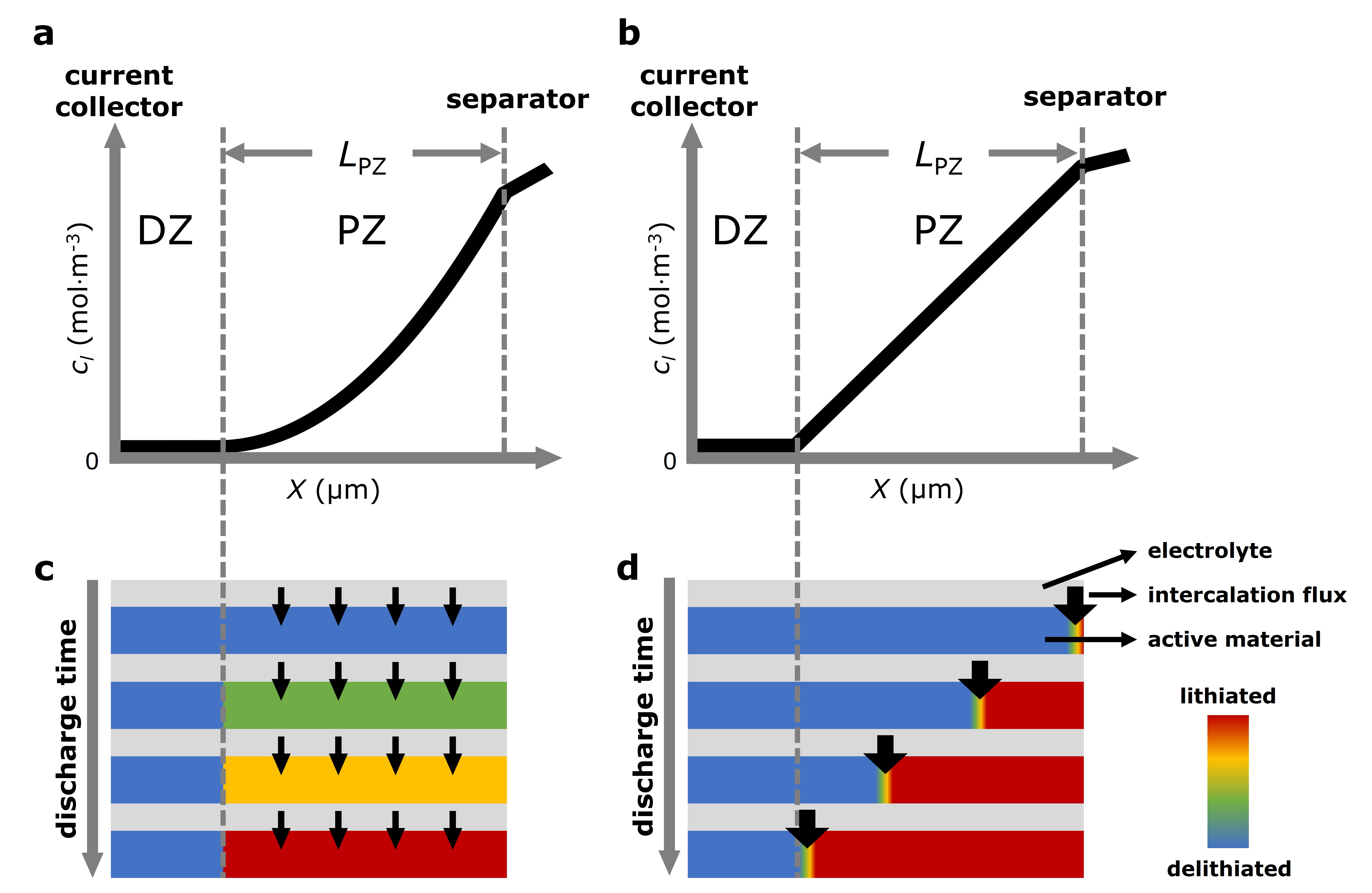}
    
    \caption{\textbf{Discharge characteristics of battery electrodes exhibiting uniform-reaction (UR) and moving-zone-reaction (MZR) behavior.}\\
    \textbf{a} Pseudo-steady-state salt distribution within an idealized UR-type half cell in the middle of discharge. $X$ denotes the distance from the cathode current collector. The cathode is separated into a salt penetration zone (PZ) and a salt depletion zone (DZ) based on salt availability in electrolyte. $L_\textrm{PZ}$ marks the size of the PZ and serves as a proxy to the (de)lithiation state. \textbf{b} Pseudo-steady-state salt distribution in an idealized MZR-type half cell toward the end of discharge.
    \textbf{c}, \textbf{d} Idealized reaction distribution in electrodes exhibiting UR and MZR behavior, respectively. The discharge process starts at time $t = 0$ and concludes at $t_{end}$.
    }\label{fig:URMZR}

\end{figure}

As a basis for deriving the UR and MZR models, Wang and Tang observed from P2D simulations  that the salt concentration $c_l$ in the electrolyte reaches a pseudo-steady-state during the discharge process. For UR-type cathodes, this pseudo-steady-state becomes apparent shortly after discharge begins. Figure \ref{fig:URMZR}a shows a typical salt concentration profile along electrode thickness direction for UR-type half cells in the middle of discharge. Salt concentration $c_l$ gradually declines from the cathode-separator interface towards the cathode current collector, reaching near-zero levels midway. This distribution delineates two distinct regions within the cathode: a salt penetration zone (PZ) where the salt concentration is non-zero and a salt depletion zone (DZ) characterized by diminishing lithium supply for electrode particle lithiation. Notably, the reaction flux remains relatively  uniform in the PZ for a significant duration of the discharge process, but minimal intercalation occurs in the DZ. As depicted in Figure \ref{fig:URMZR}c, the reaction front for an idealized UR-type battery uniformly spans the entire PZ during discharge from $t = 0$ to $t_{end}$, leaving particles in the DZ unreacted. The reaction concludes when all particles within the PZ are completely lithiated. The size of the penetration zone, denoted as $L_\textrm{PZ}$ and highlighted in the schematics, therefore indicates the degree of electrode utilization during discharge.

In a MZR-type half cell, the salt concentration profile evolves gradually during discharge.
Initially, reaction commences at the separator end and progresses toward the current collector. 
As discharge proceeds, salt from the unreacted portion of the electrode is consumed to fuel intercalation until depletion occurs in the electrolyte, marking the emergence of the pseudo-steady-state profile.
Figure \ref{fig:URMZR}b illustrates a typical salt concentration distribution towards the end of discharge for MZR-type half cells. 
In contrast to its UR counterpart, which maintains a relatively stable salt concentration distribution for much of the discharge process, the pseudo-steady-state salt concentration characteristic of MZR behavior becomes prominent only in the late stage of discharge.
The electrode reaction distribution is depicted in Figure \ref{fig:URMZR}d, where a sharp reaction front divides the electrode into a fully lithiated region (PZ) and a fully delithiated region (DZ).
At any given time $t$, only particles located at the reaction front  undergo lithium intercalation.
After these particles are fully lithiated, the intercalation flux peak shifts toward the current collector, initiating reaction in other particles.
When the reaction front propagates, salt concentration outside the lithiated region is reduced and eventually reaches zero, which marks the occurrence of complete salt depletion in the DZ and the termination of discharge. As before, electrode particles in the DZ remain fully charged and those in the PZ are fully discharged.

In full cells, reaction distribution in the anode can also be categorized as UR and/or MZR types. Graphite (Gr) anode, for example, exhibits a mixed reaction type owing to the shape of its OCP. At the onset of delithiation, graphite undergoes the LiC$_6 \rightarrow$ LiC$_{12}$ staging transition at an OCP of approximately 0.05V, demonstrating MZR behavior. After DoD rises above 30\%, graphite exhibits solid solution behavior, signifying a transition to UR type behavior. Since there is no depletion of salt in the anode during discharge, graphite particles tend to react uniformly in the entire anode when the final DoD is not too small. Consequently, we consider the graphite anode as a UR type electrode in this study. It is important to note, however, that this simplification may introduce foreseeable errors when graphite operates in the MZR regime.

\subsection{The URCs Model}\label{URCmodel}
\subsubsection{Electrolyte Transport Module}
The electrolyte transport module in the URCs model solves for the steady-state salt concentration profile in a similar way as the Wang-Tang model, but considers the more general situation where the electrolyte properties are concentration dependent and the porous electrode could have a heterogeneous structure with spatially-dependent properties, e.g. graded porosity, tortuosity and particle size. We begin with the mass conservation equation and the current continuity equation from the porous electrode theory:
\begin{align}
\epsilon_{p}(x) \frac{\partial c_l}{\partial t} = \nabla &\cdot \left[ D_{amb}(c_l)\frac{\epsilon_{p}(x)}{\tau_{p}(x)}\frac{\partial c_l}{\partial x}+\frac{(1-t_{+}(c_l)) \vec{i}(x)}{F} \right] \label{eq:masscon}\\
\nabla &\cdot \vec{i}(x) = -Fa_{p}(x)j_{in}(x)\label{eq:currentcon}
\end{align}
where $c_l$ is the salt concentration in the electrolyte and $\vec{i}(x)$ is the ionic current density. 
The above equations do not apply to the electrolyte DZ, where $c_l$ and $\vec{i}$ are assumed to be zero throughout the discharge process. 
Electrode porosity $\epsilon_p$, tortuosity $\tau_p$, and volumetric surface area $a_p = 3 (1-\epsilon_p)/r_p$ are characteristic to the specific regions of the battery cell $p\in\{cat, an, sep\}$ representing the cathode, the anode, or the separator. 

Under the assumption of steady-state electrolyte transport, Equation \ref{eq:masscon} reduces to:
\begin{equation}
\nabla \cdot \left[ D_{amb}(c_l)\frac{\epsilon_{p}(x)}{\tau_{p}(x)}\frac{\partial c_l}{\partial x} \right]=-\nabla \cdot \left[ \frac{(1-t_{+}(c_l))\vec{i}(x)}{F} \right] \label{eq:URgov}
\end{equation}
Let $x$ = 0 be at the interface between the cathode and current collector. At the PZ/DZ boundary $x = L_{cat}-L_\textrm{PZ}$, 
where $L_{cat}$ and $L_\textrm{PZ}$ are the cathode and penetration zone thickness, respectively, the salt concentration gradient and the ionic current are both zero. We can therefore integrate Equation \ref{eq:URgov} to acquire:
\begin{equation}
    D_{amb}(c_l)\frac{\epsilon_{p}(x)}{\tau_{p}(x)}\frac{\partial c_l}{\partial x} = -\frac{(1-t_{+}(c_l))\vec{i}(x)}{F}\label{eq:URfoundation}
\end{equation}
Similarly, an explicit expression of the ionic current density could be obtained from the integration of Equation \ref{eq:currentcon}:
\begin{equation}
    \vec{i}(x) = -\int\limits_{L_{cat}-L_\textrm{PZ}}^{x} F a_p(z)j_{in}(z)\,dz \label{eq:currentInt}
\end{equation}
Combining Equations \ref{eq:URfoundation} and \ref{eq:currentInt}, we couple the salt concentration with local reaction flux:
\begin{equation}
    D_{amb}(c_l)\frac{\epsilon_{p}(x)}{\tau_{p}(x)}\frac{\partial c_l}{\partial x} = (1-t_+(c_l))\int\limits_{L_{cat}-L_\textrm{PZ}}^{x} a_p(z)j_{in}(z)\,dz \label{eq:URworking}
\end{equation}
The above equation could be integrated again as:
\begin{equation}
    \int\limits_0^{c_l}\frac{D_{amb}(c_l')}{1-t_+(c_l')}\,dc_l' = \int\limits_{L_{cat}-L_\textrm{PZ}}^{x}\, dy \frac{\tau_{p}(y)}{\epsilon_{p}(y)}\int\limits_{L_{cat}-L_\textrm{PZ}}^{y} a_p(z)j_{in}(z)\,dz \label{eq:elyteEqn}
\end{equation}
If we define the left-hand side of Eq.~\ref{eq:elyteEqn} as a new function $G$:
\begin{equation}
    G(c_l) = \int\limits_0^{c_l}\frac{D_{amb}(c_l')}{1-t_+(c_l')}\,dc_l' \label{eq:fcnG}
\end{equation}
which is solely determined by the electrolyte properties, the salt concentration profile could then be expressed in terms of its inverse function: 
\begin{equation}
    c_l(x) = G^{-1}\Big(\int\limits_{L_{cat}-L_\textrm{PZ}}^{x}\, dy \frac{\tau_{p}(y)}{\epsilon_{p}(y)}\int\limits_{L_{cat}-L_\textrm{PZ}}^{y} a_p(z)j_{in}(z)\,dz\Big)\label{eq:saltCon}
\end{equation}
Under the UR assumption, the reaction flux $a_p(x)j_{in}(x)$ is proportional to the local volumetric fraction of the active materials $\nu_p(x)$. Its general expression and simplification for constant electrode porosity are given in Table \ref{tab:aj}. 

\begin{table}[!tbh]

\centering
\caption{Expression of the reaction flux $a_p j_{in}$}\label{tab:aj}%
\SetTblrInner{rowsep=10pt}
\begin{tblr}{|c|c|c|c|}
\hline
Location in Cell & \makecell{Cathode\\  ${\scriptstyle x\in [L_{cat}-L_\textrm{PZ}, L_{cat}]}$}& \makecell{Anode \\ ${\scriptstyle x\in [L_{cat}+L_{sep}, L_{cat}+L_{sep}+L_{an}]}$} & \makecell{Separator \\${\scriptstyle x\in [L_{cat}, L_{cat}+L_{sep}]}$}\\\hline
 General Expression & ${\displaystyle \frac{I\nu_{cat}(x)}{F \int_{L_{cat}-L_\textrm{PZ}}^{L_{cat}}\nu_{cat}(z)\,dz}}$ & ${\displaystyle -\frac{I\nu_{an}(x)}{F \int_{L_{cat}+L_{sep}}^{L_{cat}+L_{sep}+L_{an}}\nu_{an}(z)\,dz}}$  & 0  \\\hline
 Constant Electrode Porosity    & ${\displaystyle \frac{I}{F L_\textrm{PZ}}}$   & ${\displaystyle -\frac{I}{F L_{an}}}$  & 0  \\\hline
\end{tblr}
\end{table}

The salt concentration profile $c_l(x)$ given by Eq.~\ref{eq:saltCon} depends on the penetration zone width $L_\textrm{PZ}$, which remains unknown up to this point.
We close the loop by determining $L_\textrm{PZ}$ from the conservation of salt in the electrolyte:
\begin{align}
     \int\limits_{L_{cat}-L_\textrm{PZ}}^{L_{cat}+L_{sep}} \epsilon_p(x) c_l(x) \, dx &= c_{l0} \int\limits_{L_{cat}-L_\textrm{PZ}}^{L_{cat}+L_{sep}} \epsilon_p(x) \, dx \qquad &\mbox{(Half Cell)} \label{eq:LPZhalf}\\[1em]
    \int\limits_{L_{cat}-L_\textrm{PZ}}^{L_{cat}+L_{sep}+L_{an}} \epsilon_p(x) c_l(x) \, dx &= c_{l0} \int\limits_{L_{cat}-L_\textrm{PZ}}^{L_{cat}+L_{sep}+L_{an}} \epsilon_p(x) \, dx \qquad&\mbox{(Full Cell)} \label{eq:LPZfull}
\end{align}
where $c_{l0}$ is the average salt concentration. 

With $c_l(x)$ known, the steady-state electrolyte potential profile $\Phi_l(x)$ could be calculated  from the ionic current expression:
\begin{equation}
    \vec{i}(x) = -\frac{\epsilon_p(x)}{\tau_p(x)}  
    \left[\kappa(c_l) \frac{\partial \Phi_l}{\partial x} - \frac{2 R T (1-t_+(c_l)) \kappa(c_l)}{F c_l}\left(1+\frac{\partial \ln{f_{\pm}(c_l) }}{\partial \ln{c_l}}\right) \frac{\partial c_l}{\partial x} \right] \label{eq:ionicI}
\end{equation}
where $\kappa$ is the ionic conductivity and $\displaystyle 1+\frac{\partial \ln{f_{\pm}(c_l) }}{\partial \ln{c_l}}$ is the thermodynamic factor of the electrolyte. Using Equation \ref{eq:URfoundation} to eliminate $dc_l/dx$ in Equation $\ref{eq:ionicI}$, we obtain:
\begin{equation}
    \frac{\partial \Phi_l}{\partial x} = -\frac{\tau_p(x) \omega(c_l)}{\epsilon_p(x) \kappa(c_l)} \vec{i}(x) \label{eq:dPhil}
\end{equation}
with 
\begin{equation}
    \omega(c_l)\equiv 1+\frac{2 R T \kappa(c_l) (1+\frac{\partial \ln{f_{\pm}(c_l) }}{\partial \ln{c_l}}) (1-t_+(c_l))^2}{F^2 c_l D_{amb}(c_l)}
\end{equation}
$\Phi_l$ could be solved by replacing $\vec{i}$ with Eq.~\ref{eq:currentInt} in Eq.~\ref{eq:dPhil} and then integrating the equation:
\begin{equation}
    \Phi_l(x) = \int\limits_{L_{cat}+L_{sep}}^{x}\frac{\tau_p(y) \omega(c_l(y))\,dy}{\epsilon_p(y) \kappa(c_l(y))} \int\limits_{L_{cat}-L_\textrm{PZ}}^{y} F a_p(z) j_{in}(z) dz \label{eq:phil}
\end{equation}
Note that $\Phi_l$ at the separator/anode interface $x=L_{cat}+L_{sep}$ is set to be zero here.

\subsubsection{Solid Phase Module}
In the previous Wang-Tang model~\autocite{wang_quantitative_2020}, all the active material particles in the PZ are assumed to be fully reacted at the end of discharge. The normalized discharge capacity is thus given by $\mathrm{DOD_f}=L_\textrm{PZ}/L_{cat}$. 
In the URCs model, the electrode particles in the PZ may have $<$100\% depth of discharge (DoD) due to slow solid diffusion while the particles in the DZ are still assumed to be fully non-reacted.
For reason to be discussed below, we also do not assume the electrode particles in the PZ to have the same DoD.
According to the Butler-Volmer equation, the reaction flux $j_{in}$ is controlled by the overpotential $\eta$, which is given by:
\begin{equation}
    \eta = \Phi_s(x) - \Phi_l(x) - U_{eq}(c_{ss})
    \label{eq:etaEqn}
\end{equation}
where $U_{eq}$ is the local OCP of the electrode particles, which is a function of the lithium surface concentration $c_{ss}$. 
Because $\Phi_l(x)$ and $\Phi_s(x)$ vary spatially,      
$U_{eq}$ must have a gradient within the electrode when a uniform $j_{in}$ or $\eta$ develops.
For UR-type electrodes, this means the presence of a $c_{ss}$ gradient as their $U_{eq}$ is sensitive to lithium stoichiometry.
Therefore, uniform reaction is not the same as uniform SoC.
A transient period must precede the UR behavior during discharge to establish such an SoC gradient. 
The Wang-Tang model neglects this transient period and assumes that UR occurs throughout the discharge process so that all the particles reach 100\% DoD simultaneously.
Here we remove this approximation and allow the particle-level SoC (or DoD) to be non-uniform within the PZ in order to further improve the prediction accuracy.

In the solid phase module, we first determine the spatial distribution of $c_{ss}$ within the PZ from Eq.~\ref{eq:etaEqn}, and then correlate $c_{ss}$ to the particle-level DoD through the solution to the solid-diffusion equation. The electrode-level DoD as a function of the cell voltage is then obtained by integration. Using the expression of $j_{in}$ given in Table~\ref{tab:aj}, the overpotential $\eta(x)$ is solved from the Butler-Volmer equation:
\begin{equation}
    \eta = -\frac{2 R T}{F} \sinh^{-1}(\frac{Fj_{in}}{2i_0})\label{eq:BV}
\end{equation}
where the anodic/cathodic transfer coefficients are assumed to be $1/2$. The exchange current density $i_0$ is expressed as:
\begin{equation}
    i_0(x) = F k_0\sqrt{c_l(x) c_{ss} (c_{smax}-c_{ss}) }
    \label{eq:j0}
\end{equation}
where $k_0$ is the reaction rate constant. 
As an approximation, $c_{ss}$ is replaced with $(c_{smax, cat} + c_{s0, cat})/2$ for the cathode and $c_{s0, an}/2$ for the anode in full cell.
For simplicity, we shall also assume that the electrical conductivity in the solid phase is suffiently high so that the solid potential is uniform across the cathode and anode: $\Phi_s=\Phi_{s,cat/an}$, where $\Phi_{s,cat/an}$ represent the cathode/anode terminal potentials.    
This assumption could be easily relaxed. 

By substituting Eqs.~\ref{eq:phil} and \ref{eq:BV} into Eq.~\ref{eq:etaEqn}, we could solve for the $c_{ss}$ profile within the PZ in the cathode and also the anode in the case of full cell:
\begin{align}\label{eq:cssvphis}
    c_{ss, cat}(x,\Phi_{s,cat}) &= U_{eq, cat}^{-1}\big(\Phi_{s,cat} - \Phi_l(x) - \eta_{cat}(x) \big) \\[1em]
    c_{ss, an}(x, \Phi_{s,an}) &= U_{eq, an}^{-1}\big(\Phi_{s,an} - \Phi_l(x)-\eta_{an}(x) \big)\tag{Full Cell Only}
\end{align}
where $U_{eq}^{-1}$ is the inverse function of $U_{eq}(c)$. The $c_{ss}$ profile evolves with the terminal potentials during the discharge process. When solid diffusion is facile, $c_{ss}$ provides a good approximation to the average Li concentration in the particle $\overline{c}_s$. However, here we also need to consider the situation where $c_{ss}$ differs significantly from $\overline{c}_s$ due to sluggish diffusion, i.e. when $r^2_p/D_{s,p}$ is much larger than the discharge time. To determine $\overline{c}_s$, we solve the solid diffusion equation in a spherical particle:
\begin{equation}\label{eq:sphDiffusion}
    \frac{\partial c_s}{\partial t} = \frac{D_{s, p}}{r^2} \frac{\partial}{\partial r}
    \left(r^2\frac{\partial c_s}{\partial r}\right) 
\end{equation}
with the constant flux boundary condition $-D_{s,p}\partial c_s/\partial r|_{r=r_p}=j_{in}$ and initial condition $c_s(r,t=0)=c_{s0,p}$. An analytical solution to the equation exists\autocite{carslaw_conduction_1986}. In particular, the surface Li concentration has the following expression:
\begin{equation}\label{eq:css}
    c^{sol}_{ss, p}(t; j_{in}) = c_{s0, p} + j_{in}\bigg[\frac{3 t}{r_p}+\frac{r_p}{5 D_{s, p}}-\frac{2 r_p}{D_{s, p}}\sum_{m = 1}^{\infty}\lambda_{m}^{-2}\exp\big(-\frac{\lambda_m^2 D_{s, p} t}{r_p^2}\big)\bigg]
\end{equation}
where $\lambda_m$ is the $m-$th positive root of the equation $\tan(\lambda) = \lambda$. 
Eq.~\ref{eq:css} shows that $c_{ss}$ gradually deviates from $c_{s0}$ during discharge. We use the time it takes for $c_{ss}$ to reach the value specified in Eq.~\ref{eq:cssvphis} to estimate the amount of Li (de)intercalated into an individual particle:
\begin{equation} \label{eq:csavg}
    \Delta{c}^{sol}_{s,p}(x, \Phi_{s,p}) = \frac{3 j_{in} T_p(c_{ss,p}(x, \Phi_{s,p}); j_{in})}{r_p}
\end{equation}
where $T_p(c_{ss};j_{in})$ is the inverse function of $c^{sol}_{ss,p}(t; j_{in})$ in Eq.~\ref{eq:css}. 

The electrode-level DoD could be obtained through the integration of $\Delta{c}^{sol}_{s,p}$:
\begin{align}
    \textrm{DoD}_{cat}(\Phi_{s,cat}) &= \frac{\int_{L_{cat}-L_\textrm{PZ}}^{L_{cat}} \Delta{c}^{sol}_{s, cat}(x, \Phi_{s,cat}) \nu_{cat}(x) \,dx}{(c_{smax,cat}-c_{s0,cat})\int_0^{L_{cat}} \nu_{cat}(x) \, dx}\label{eq:DoDcat}\\[1em]
    \textrm{DoD}_{an}(\Phi_{s,an}) &= \frac{\int_{L_{cat}+L_{sep}}^{L_{cat}+L_{sep}+L_{an}} \Delta{c}^{sol}_{s, an}(x, \Phi_{s,an}) \nu_{an}(x) \,dx}{c_{s0,an} \int_0^{L_{cat}} \nu_{an}(x)\, dx}\tag{Full Cell Only}
\end{align}
The inverse functions of Eq.~\ref{eq:DoDcat}, $\Phi_{s,cat}(\textrm{DoD})$ and $\Phi_{s,an}(\textrm{DoD})$, relate the terminal potentials at the current collectors with DoD. Using them, the cell voltage $U_\textrm{out}$ could be expressed as a function of DoD, which is:
\begin{align}
    U_\textrm{out}(\textrm{DoD}) & = \Phi_{s,cat}(\textrm{DoD}) - \eta_{Li}\qquad&\mbox{(Hall Cell)} \label{eq:UoutHalf} \\[1em]
    U_\textrm{out}(\textrm{DoD}) &= \Phi_{s,cat}(\textrm{DoD}) - \Phi_{s,an}(\textrm{DoD})\qquad&\mbox{(Full Cell)} \label{eq:UoutFull}
\end{align}
In Eq.~\ref{eq:UoutHalf}, $\eta_{Li}$ is the overpotential at the Li metal anode surface and given by $2RT\sinh^{-1}(I/(2i_0^{Li}))/F$. 
Eq.~\ref{eq:UoutHalf} or \ref{eq:UoutFull} predicts the discharge voltage curve and also the final DoD (DoD\textsubscript{f}) upon reaching the cutoff voltage $U_{\mathit{cutoff}}$:
\begin{equation}
    \mathrm{DoD_f} = U_\textrm{out}^{-1}(U_{\mathit{cutoff}})
    \label{eq:DoDf}
\end{equation}

We briefly summarize the key steps in the development of the URCs model. We start by employing the steady-state electrolyte transport and UR assumptions to determine the electrolyte PZ width $L_\textrm{PZ}$ (Eq.~\ref{eq:LPZhalf} or \ref{eq:LPZfull}) and the distributions of salt concentration $c_l(x)$ (Eq.~\ref{eq:saltCon}) and electrolyte potential $\Phi_l(x)$ (Eq.~\ref{eq:phil}) within the PZ.
The calculated $\Phi_l(x)$ is used to determine the OCP of the cathode/anode material  (Eq.~\ref{eq:etaEqn}) at a given terminal potential, from which the surface lithium concentration in electrode particles $c_{ss}$ is obtained (Eq.~\ref{eq:cssvphis}). The amount of Li (de)intercalated in the particles $\Delta c^{sol}_{s,p}$ is then estimated from $c_{ss}$ (Eq.~\ref{eq:csavg}) based on the solution to the solid diffusion equation. Finally, the integration of particle-level DoD within the PZ allow us to predict the discharge voltage curve (Eq.~\ref{eq:UoutHalf} and \ref{eq:UoutFull}) and the discharge capacity (Eq.~\ref{eq:DoDf}).

The URCs model derived above has been implemented in MATLAB. The open source code is available online (see Data Availability).

\subsection{Comparison with P2D Simulations}\label{sec:Figs}
The URCs model is benchmarked against P2D simulations in a series of comparative studies. Tests are conducted for both NMC half cells and NMC/Gr full cells with the cathode porosity fixed at 0.25. In full cell configurations, the anode to cathode thickness ratio is chosen to be $L_{an}:L_{cat} = 1.15$ and the anode porosity $\epsilon_{an}$ is set to fix the anode to cathode capacity ratio at 1.1. Cutoff voltage is 3.0V for NMC half cells and 2.8V for NMC/Gr full cells. The electrolyte properties used in this study are based on a LiPF\textsubscript{6} in PC/EC/DMC electrolyte reported by Valøen and Reimers\autocite{valoen_transport_2005}. Electrode tortuosity relationships are derived from the study on calendared electrodes conducted by Usseglio-Viretta et al.\autocite{usseglio-viretta_resolving_2018}. The parameter values used in the URCs and P2D calculations are listed in Table \ref{Stab:P2Dparams}. 

\subsubsection{Mass and potential distributions}
\begin{figure}[!tbh]
    \centerline{
    \includegraphics[width = 1.1\textwidth]{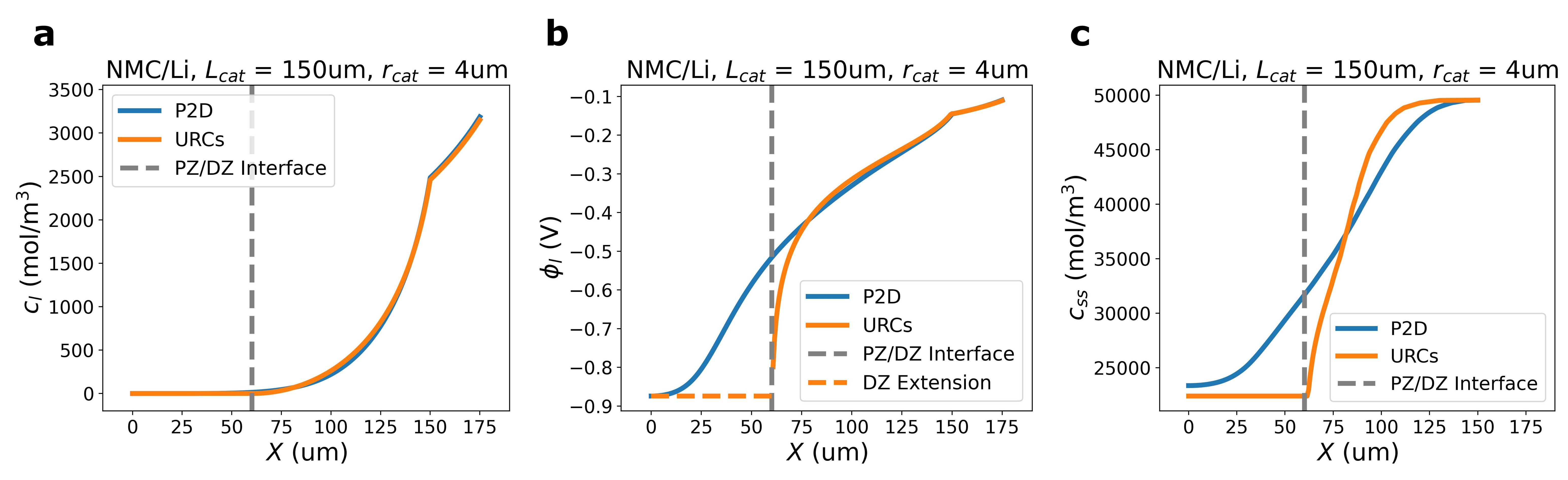}
    }
    \caption{\textbf{Comparison of mass and potential distributions predicted by the URCs model and P2D simulation.}\\
     Results from the URCs model and the P2D simulation for an NMC half cell with a cathode thickness $L_{cat} = 150\mu$m and cathode particle size $r_{cat} = 4\mu$m discharged to DoD = 45\% at 2C rate: \textbf{a} salt concentration $c_l$, \textbf{b} electrolyte potential $\Phi_l$ and \textbf{c} surface lithium concentration in NMC particles $c_{ss}$. 
     The PZ/DZ interface predicted by the URCs model is marked by the vertical dashed line. $X$ is the distance from the cathode current collector.
    }\label{fig:varcomps}
   
\end{figure}

First, we compare the mass and potential distributions in NMC half cells as solved by the two approaches. Figure \ref{fig:varcomps} shows $c_l$, $\Phi_l$ and $c_{ss}$ from P2D simulations versus the URCs model for an NMC half cell undergoing 2C discharge.
The P2D simulation results are taken from an intermediate state at DoD $\approx$ 45\%, at which the UR behavior has been established. 
The URCs model predicts a PZ width $L_\textrm{PZ} \approx 90\mu$m, placing the PZ/DZ boundary at a distance of $\approx 60\mu$m from the current collector ($X$=0). 
As shown in Figure \ref{fig:varcomps}a, the predicted $c_l$ from the URCs model agrees very well with P2D. This validates the steady-steady transport and the UR assumptions that are used to simplify the electrolyte mass conservation equation from the porous electrode theory. 

Figure \ref{fig:varcomps}b shows that $\Phi_l$ calculated by the URCs and P2D have good agreement inside the PZ. 
However, $\Phi_l$ diverges to infinity at the PZ/DZ boundary in the URCs model because the salt is assumed to be fully depleted in the DZ. In contrast, $\Phi_l$ exhibits a more gradual transition and extends into the DZ in P2D simulation. 
Such difference is caused by the fact that $c_l$ in P2D is not strictly zero within the DZ and a small reaction flux thus still exists in this region. 
To handle $\Phi_l$ properly in the calculation, we truncate it below a cutoff value $\Phi^{\mathrm{cutoff}}_l$ (horizontal dashed line in Figure \ref{fig:varcomps}b), i.e. replacing $\Phi_l$ with $\max(\Phi_l, \Phi^{\mathrm{cutoff}}_l)$ in the URCs model. Though the value of $\Phi^{\mathrm{cutoff}}_l$ is chosen somewhat arbitrarily, tests show that it has little effect on the URCs predictions of the discharge voltage curve and capacity. 

Compared to $c_l$ and $\Phi_l$, $c_{ss}$ predicted by URCs shows more noticeable deviation from the P2D result, see Figure \ref{fig:varcomps}c. Unlike the URCs prediction, cathode particles are partially lithiated inside the DZ in the P2D simulation due to the nonvanishing reaction flux. On the other hand, they are less lithiated within the PZ compared to URCs. 
This relatively large deviation could be attributed to the errors accumulated in the evaluation of $\Phi_l$ and $\eta$, which are used to calculate the distribution of $U_{eq}$ and then $c_{ss}$ across the PZ.
Besides the error associated with $\Phi_l$, the UR assumption also introduces inaccuracy in the calculation of $\eta$. Additionally, the exchange current density is approximated as independent of $c_{ss}$, which is another source of error.
However, as we will demonstrate below, the errors inherent in the predicted $c_{ss}$ distribution are offset upon summation, which leads to better estimates of the discharge capacity and voltage curve.

\subsubsection{Discharge Capacity}
\begin{figure}[H]
    \centering
    \includegraphics[width = \textwidth]{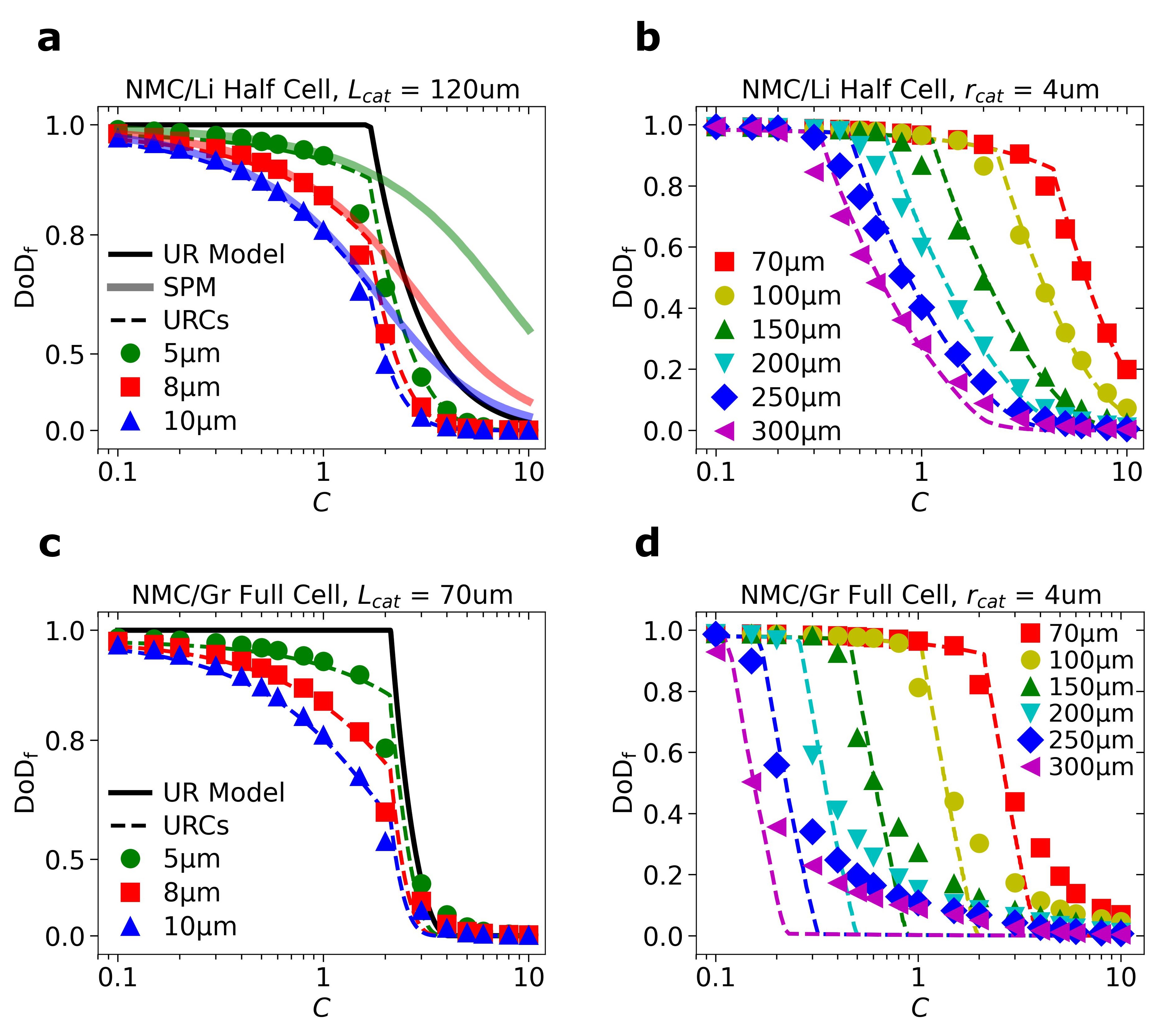}
    \caption{\textbf{Predicting discharge rate performance for NMC half cells and NMC/Gr full cells with P2D simulations, the URCs model and other simplified models.}\\
        \textbf{a} The advantage of regulating mixed reaction kinetics within the URCs model (dashed lines) is highlighted through comparison with the single particle model (solid colored lines) and the UR model (solid black line) on an NMC half cell with fixed $L_{cat}=120\mu\textrm{m}$ and varying $r_{cat}$. The URCs model agrees well with the P2D simulations (colored symbols). SPM overlooks the pronounced salt depletion effect at C-rates above $C_\textrm{crit}$. The UR model incorrectly assumes facile solid diffusion regardless of particle size. The vertical axis is scaled to offer a better view of the particle size effect at larger DoD\textsubscript{f}. \textbf{b} The agreement between the discharge capacities predicted by URCs and P2D is retained in NMC half cells across a wide range of $L_{cat}$ and C-rates. 
        \textbf{c} Normalized discharge capacities predicted by the P2D simulation (colored symbols), UR (solid line) and URCs (dashed lines) models for NMC/Gr full cells with $L_{cat}$ = 70$\mu$m and different $r_{cat}$.
        \textbf{d} Normalized discharge capacities predicted by P2D (colored symbols) and URCs (dashed lines) for NMC/Gr full cells with $r_{cat}$ = 4 $\mu$m and different cathode thickness.  
    }\label{fig:LC}
    
\end{figure}

Next we examine the performance of the URCs model in predicting the discharge rate capability of battery cells. 
In Figure \ref{fig:LC}a, we plot the predictions by the URCs model, the single particle model (SPM) and the original Wang-Tang's UR model against P2D simulations for NMC half cells with $L_{cat}$ = 120 $\mu$m and variable particle radius (5 -- 10 $\mu$m). The UR model (solid black line) predicts a critical C rate $C_\textrm{crit}$, at which salt depletion occurs in the electrolyte. Below $C_\textrm{crit}$, the PZ spans the entire cathode and the predicted DoD\textsubscript{f} is always 1 because the electrode particles are assumed to be fully lithiated. Above $C_\textrm{crit}$, DoD\textsubscript{f} decreases below 1 as the PZ width is reduced by sluggish electrolyte transport. On the other hand, the SPM (solid colored lines) considers solid diffusion in active material particles to be the only rate-limiting reaction mechanism. Transport in the electrolyte is assumed to be sufficiently facile and therefore free from salt concentration or potential gradient. As the size of the active material particles increases, solid diffusion becomes slower due to the longer diffusional pathway, resulting in reduced rate performance predicted by the SPM. Figure \ref{fig:LC}a clearly shows that the UR model overestimates the discharge capacity when the particle size is large and solid diffusion is no longer facile. The SPM also significantly overestimates the rate performance above $C_\textrm{crit}$, where salt depletion becomes the predominant factor behind the deteriorating cell performance. In contrast, the URCs model (dashed lines) exhibits excellent agreement with P2D at all the tested particle radii. In particular, the URCs model captures the gradual decrease of DoD\textsubscript{f} with the C rate below $C_\textrm{crit}$, which is caused by solid diffusion alone. It also predicts with good accuracy the precipitous drop of DoD\textsubscript{f} above $C_\textrm{crit}$, which is inflicted jointly by limitations in electrolyte transport and solid diffusion. We note that $C_\textrm{crit}$ is the same in both the URCs and the UR models because the URCs use the same criteria to predict the onset of salt depletion in electrolyte. Similarly good agreement between URCs and P2D is also seen in NMC/Gr full cells as shown in Figure \ref{fig:LC}c. The above test confirms that the URCs model combines the UR and SPM models as intended and is able to accurately predict the discharge behavior of battery cells under mixed kinetic control.  

As a further test, we compare the URCs predictions with the P2D simulations over a wide range of the electrode thickness (70 -- 300 $\mu$m) in half and full cells in Figure \ref{fig:LC}b and \ref{fig:LC}d, respectively. A high-level of agreement is seen in half cell configurations and when DoD\textsubscript{f} $>30$\% for full cells. At lower DoD\textsubscript{f}, however, the URCs model tends to underestimate the discharge capacity of the full cells, especially when electrodes are thick. Such discrepancy is also observed in the comparison between the UR model and P2D simulations\autocite{wang_quantitative_2020}. It is due to the hybrid reaction behavior exhibited by the graphite anode. While the URCs model assumes graphite to behave as an UR-type electrode, the OCP curve of graphite consists of both sloped and flat segments, which correspond to the solid solution and two-phase coexistence regions, respectively. During discharge, the lithiated graphite anode goes through the I-II staging transition at DoD$<$ 30\%, in which it displays the MZR behavior. This explains the relatively large error of the URCs model in the low DoD\textsubscript{f} regime. As the discharge process progresses to higher DoD, the graphite's OCP curve enters a more sloped region. Its reaction behavior accordingly becomes more UR-like and the agreement between the URCs and P2D thus improves. When applying the URCs model to optimizing the full cell configuration, its relatively poor prediction accuracy at low DoD\textsubscript{f} is not a practical  concern because the optimization objective is to identify cell parameters that achieve high DoD\textsubscript{f}, where the URCs model performs well.

\subsubsection{Discharge Voltage Curves and Energy Output}
\begin{figure}[H]
    \centering
    \includegraphics[width = 0.9\textwidth]{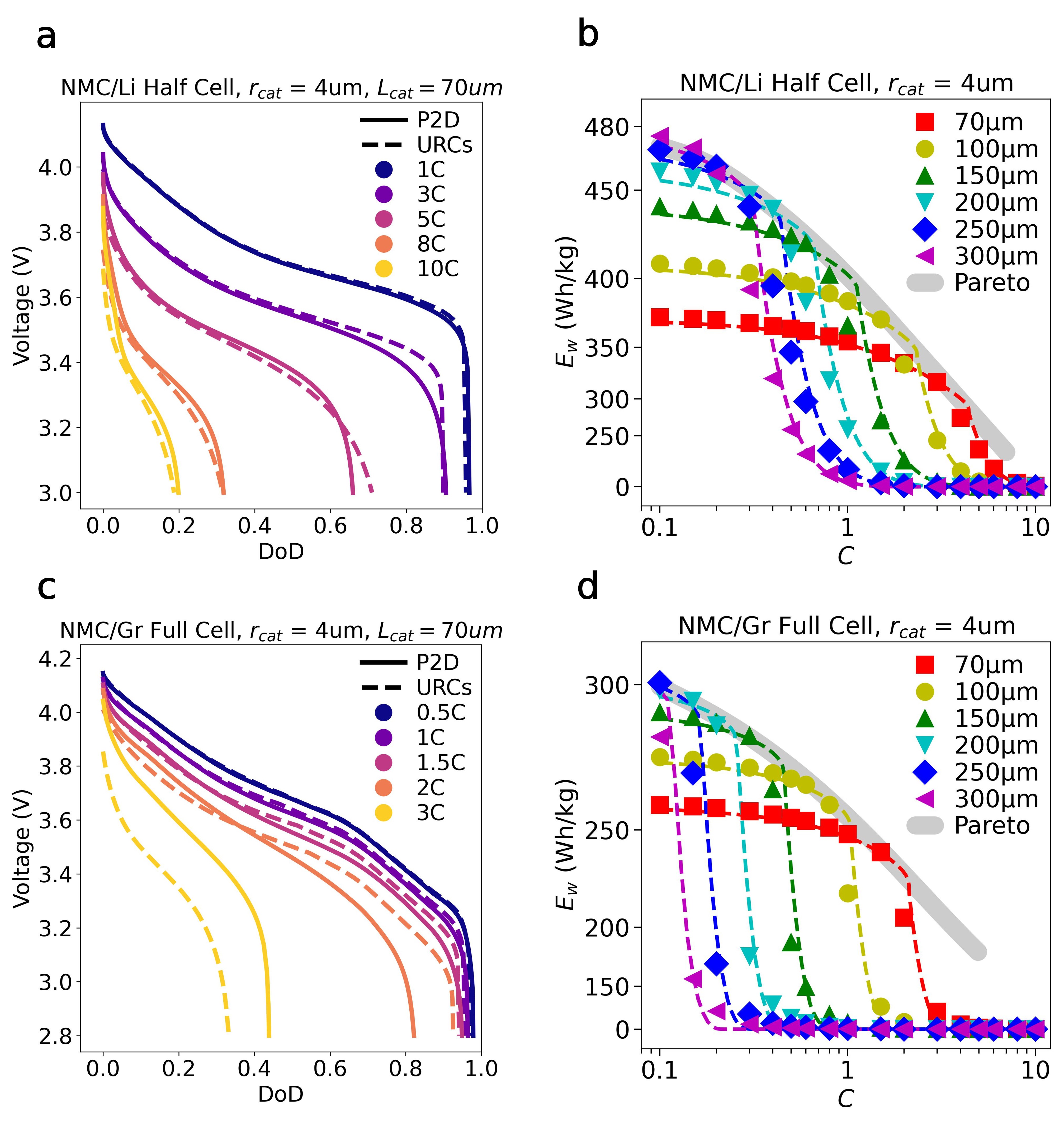}
    \caption{\textbf{Predicting discharge voltage curves and specific energy output of NMC half cells and NMC/Gr full cells with the URCs model.}\\
    \textbf{a} URCs-predicted discharge voltage curves (dashed lines) vs P2D results (solid lines) for an NMC half cell with $L_{cat} = 70\mu\textrm{m}$ and $r_{cat} = 4\mu\textrm{m}$. \textbf{b} URCs-predicted cell-level specific energy $E_w$ (dashed lines) in comparison to P2D simulations (colored symbols) for NMC half cells with $r_{cat}$ = 4 $\mu$m and different cathode thickness. 
    The URCs model reveals the Pareto front (solid line) that facilitates battery system design decisions involving trade-offs between energy and power outputs. The vertical axis is scaled to offer a better view of the upper range of $E_w$ where crossovers occur. 
    \textbf{c} Discharge voltage curves predicted by URCs (dashed lines) and P2D (solid lines) for an NMC/Gr full cell with $L_{cat} = 70\mu\textrm{m}$ and $r_{cat} = 4\mu\textrm{m}$.
    \textbf{d} $E_w$ predicted by URCs (dashed lines) versus P2D (colored symbols) for NMC/Gr cells with $r_{cat}$ = 4 $\mu$m and different cathode thickness. 
    }\label{fig:Energy}
    
\end{figure}

Prediction of the voltage curves upon galvanostatic discharging based on Eq.~\ref{eq:UoutHalf} and \ref{eq:UoutFull} is an added capability of the URCs model. In Figures \ref{fig:Energy}a and \ref{fig:Energy}c, we compare the voltage curve predictions between P2D simulations and the URCs model for a half and full cell with a cathode thickness $L_{cat}$ of 70$\mu$m at several C rates. The agreement between the two is very good for the half cell, even at high rates up to 10C. The qualitative features of the voltage curves are well captured by the URCs model.
For the full cell, the agreement is less satisfactory at relatively high rates, where DoD\textsubscript{f} is low. 
The underlying reason is similar to the trend seen in the discharge capacity prediction and caused by the hybrid reaction behavior of the graphite anode. 

The areal discharge energy of the battery cell is given by the area underneath the discharge voltage curve: $\displaystyle E_A = Q_0 \int_{0}^{\mathrm{DoD_f}} V d\mathrm{DoD}$ where $Q_0$ is the areal capacity of the cathode. In Figure \ref{fig:Energy}b and \ref{fig:Energy}d, we compare the cell-level specific energy $E_w$ predicted by the URCs model, which is calculated from $E_A$ using the cell parameters in Table \ref{Stab:QwParams}, with the P2D results for NMC/Li and NMC/Gr cells across a range of cathode thickness. Overall, the URCs-predicted discharge energy displays a comparable level of agreement with P2D as the discharge capacity shown in Figure \ref{fig:LC}b and \ref{fig:LC}d. 

Figure \ref{fig:Energy}b and \ref{fig:Energy}d are analogous to the Ragone plot\autocite{ragone_review_1968} and reveal the classic trade-off between the energy and power outputs of battery cells. The $E_w$ vs C rate curves display a typical ``knee point'', beyond which salt depletion occurs in the electrolyte and the discharge energy drops precipitously. While cells with thicker electrodes boast higher maximum $E_w$, thanks to a smaller weight fraction of inactive components in the cells, their discharge energies decrease more rapidly with the discharge rate and exhibit a crossover with those with thinner electrodes, resulting in inferior performance under high power demand. The envelop of these curves, or the Pareto front, inform the highest achievable $E_w$ at a given C rate and the corresponding electrode thickness. We see that the URCs model is able to accurately predict the Pareto front and thus guide the selection of the optimal electrode thickness for a given power requirement with other cell parameters fixed. In the following sections, we apply the URCs model to battery cell optimization in more complex scenarios where two or more cell parameters are adjustable.

\subsection{Optimizing Battery Cells with the URCs Model}
\subsubsection{Grid-space Search}\label{sec:grid-space search}
In this section, we set out to optimize battery cell parameters with the URCs model and benchmark its accuracy and efficiency against the P2D simulations. As a first test, we use grid-space search to maximize the cell-level specific capacity $Q_w$ or specific energy $E_w$ at 1C discharge against cathode thickness $L_{cat}$ and porosity $\epsilon_{cat}$ for NMC/Li and NMC/Gr cells. The cathode particle radius is assumed to be 4$\mu$m. As in the previous section, the $L_{an}$:$L_{cat}$ ratio is fixed at 1.15 in NMC/Gr full cells, and anode porosity $\epsilon_{an}$ is set to maintain an anode to cathode capacity ratio of 1.1. Other cell component information are listed in Table \ref{Stab:QwParams}. 

When using P2D simulation for the optimization, we conduct the search with a two-level resolution because of its high computational cost. The $L_{cat}$--$\epsilon_{cat}$ space is first scanned on a coarse grid, which samples 15 values of $L_{cat}$ between 50$\mu$m and 400$\mu$m with a increment of 25$\mu$m and 19 values of $\epsilon_{cat}$ between 0.15 and 0.60 with an increment of 0.025. Subsequently, a refined search is performed out on a finer grid in the neighborhood of the ($L_{cat}$, $\epsilon_{cat}$) grid point that maximizes $Q_w$ or $E_w$. The search region is within $\pm 25\mu$m and $\pm 0.05$ around the approximate optimum $L_{cat}$ and $\epsilon_{cat}$, respectively, and the grid spacings are (2$\mu$m, 0.005).
We use this fine scan to identify the global optimal cathode thickness $L_{cat}^\textrm{opt}$ and porosity $\epsilon_{cat}^\textrm{opt}$. A total of 831 simulations (285 coarse scan and 546 refined scan) are carried out. 
When using the URCs model in the grid-space search, we scan the same region of the parameter space with higher resolution thanks to the model's computation efficiency. 1000 uniformly spaced values of $L_{cat}$ and $\epsilon_{cat}$ each are sampled in a total of $10^6$ calculations. The optimal cell configuration predicted by the URCs model is denoted as $L_{cat}^\textrm{opt*}$ and $\epsilon_{cat}^\textrm{opt*}$.

\begin{figure}[!tbh]
    \centerline{
    \includegraphics[width = 1.1\textwidth]{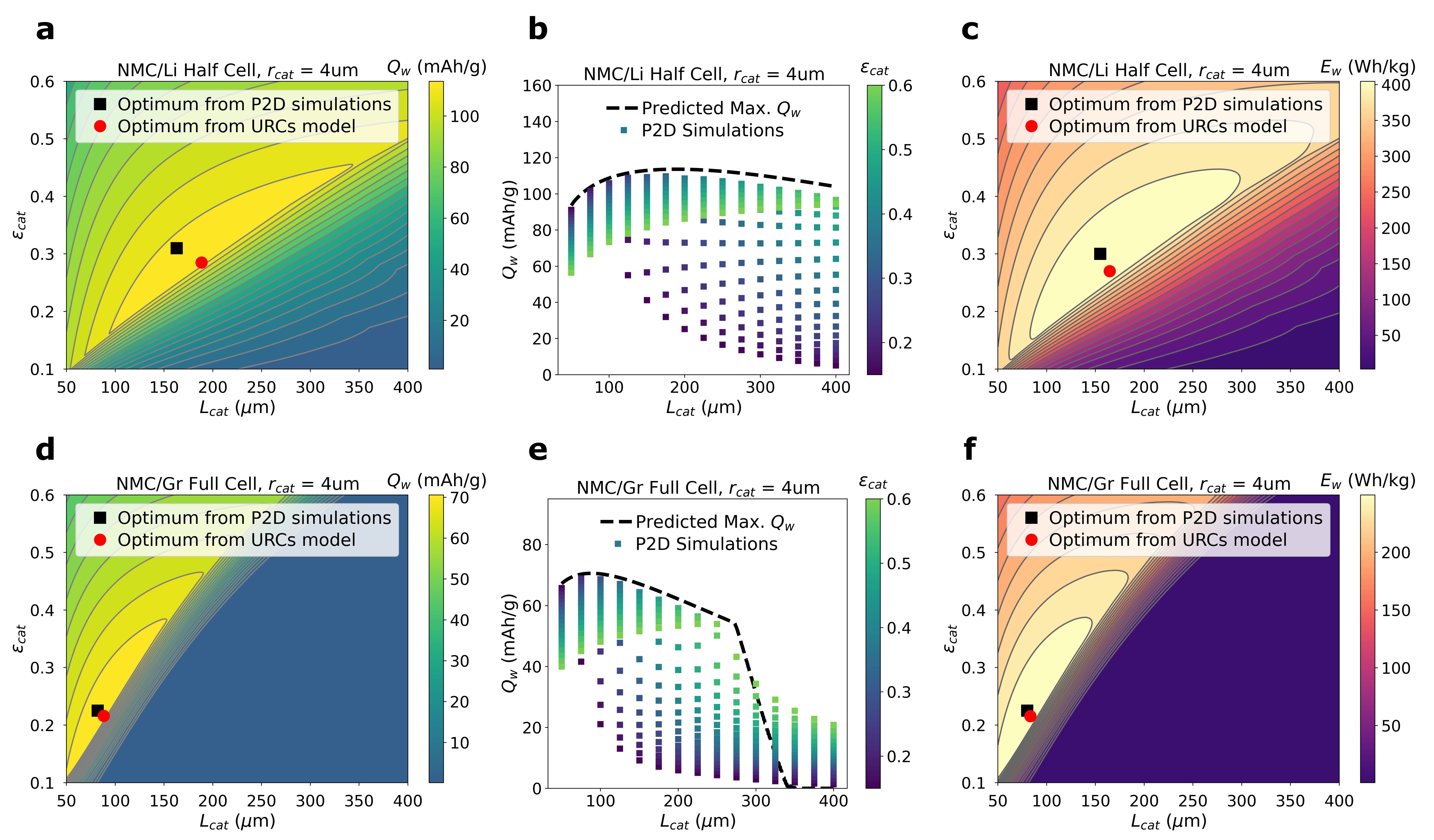}
    }

    \caption{\textbf{Optimization of NMC/Li and NMC/Gr full cells to maximize the cell level specific capacity $Q_w$ and specific energy $E_w$ at 1C discharge with grid-space search.}\\
    \textbf{a} Global optimal $Q_w$ for an NMC half cell determined from two-step grid-space search with P2D simulations (black square) and high-resolution grid-space scan with the URCs model (red circle), overlaid on URCs-generated contour plot on the $L_{cat}$-$\epsilon_{cat}$ plane. \textbf{b} The URCs model correctly predicts the upper limit of $Q_w$ (dashed envelope) achievable at each $L_{cat}$. P2D simulations with varying $\epsilon_{cat}$ are indicated by colored square symbols. \textbf{c} Global optima for $E_w$ on the $L_{cat}$-$\epsilon_{cat}$ plane of an NMC half cell, determined by P2D (black square) and URCs (red circle) using the same grid-space search strategy as for $Q_w$ optimization. \textbf{d}, \textbf{e}, and \textbf{f} are analogous to \textbf{a}, \textbf{b}, and \textbf{c} but for NMC/Gr full cells. 
    }\label{fig:GridSpaceSearch}

\end{figure}

Figure \ref{fig:GridSpaceSearch}a and \ref{fig:GridSpaceSearch}d present the contour plots of the URCs-generated $Q_w$ from the grid-space search and also mark the locations of the optimal cell configurations identified by P2D (black square) and URCs (red circle). For NMC half cells, the optimal cell configuration from P2D simulations is $(L_{cat}^\textrm{opt}, \epsilon_{cat}^\textrm{opt}) = (163\mu\textrm{m}, 0.31)$, and $Q^\textrm{opt}_w$ = 109.8 mAh/g. The URCs model predicts $(L_{cat}^\textrm{opt*}, \epsilon_{cat}^\textrm{opt*}) = (187.9\mu\textrm{m}, 0.285)$ and $Q^\textrm{opt*}_w$ = 113.6 mAh/g, which deviate from the P2D results by $15.2\%$, $8.2\%$ and $3.5\%$, respectively. For NMC/Gr full cells, we obtain $(L_{cat}^\textrm{opt}, \epsilon_{cat}^\textrm{opt}) = (82\mu\textrm{m}, 0.225)$ and $Q^\textrm{opt}_w$ = 69.1 mAh/g from P2D and $(L_{cat}^\textrm{opt*}, \epsilon_{cat}^\textrm{opt*})=(88.2\mu\textrm{m}, 0.216)$ and $Q^\textrm{opt*}_w$ = 70.6 mAh/g from URCs. The relative differences are less than $7.6\%$,  $4.0\%$ and $2.2\%$, respectively. The optimal parameters found by URCs show good agreement with the P2D simulations, with the half cell predictions having slightly larger errors than full cells. 
In Figure \ref{fig:GridSpaceSearch}b and \ref{fig:GridSpaceSearch}e, we visualize the P2D calculations of $Q_w$ from the coarse grid search by grouping them according to cathode thickness $L_{cat}$ for half and full cells, respectively. 
Additionally, the maximum obtainable $Q_w$ for each $L_{cat}$ is calculated from the URCs model and plotted as a dashed line. 
It can be seen that the URCs model well captures the upper limit of $Q_w$ for both types of cells, with the exception of full cell configurations with large $L_{cat}$, for which the URCs model underestimates $Q_w$.
It is noticeable that the upper envelope of $Q_w$ is flatter and varies more gradually with $L_{cat}$ for half cells than for full cells. This explains why the URCs model has a relatively large error in $L^\textrm{opt}_{cat}$ for half cells as their $Q_w$ is not as sensitive to $L_{cat}$ as in full cells, which is also reflected by the sparser contour lines of the half cells (Figure \ref{fig:GridSpaceSearch}a) compared with the full cells (Figure \ref{fig:GridSpaceSearch}d).

Figure \ref{fig:GridSpaceSearch}c and \ref{fig:GridSpaceSearch}f show the URCs calculations of $E_w$ as a function of $L_{cat}$ and $\epsilon_{cat}$ for half and full cells, respectively. The optimal cell configurations that maximize $E_w$ as determined by URCs and P2D are also identified in the plots. Compared to $Q_w$, the URCs model performs even better when used to optimize against $E_w$. For half cells, it differs from P2D by $6.2\%$, $9.9\%$ and $1.3\%$ in the predictions of $L_{cat}$, $\epsilon_{cat}$ and $E^{opt}_w$, respectively, and the relative differences further reduce to $4.1\%$, $4.2\%$ and $1.2\%$ for full cells. The cathode thickness optimized for $E_w$ is slightly lower than that optimized for $Q_w$. This is because increasing electrode thickness not only reduces the salt PZ width but also increases the cell resistance, which pushes the maximum $E_w$ towards lower $L_{cat}$.

While the URCs model could be used to find optimal battery configurations with good accuracy, its computation efficiency is far more superior than the P2D simulation. 
We benchmarked their performance on a Windows laptop (2.3 GHz Intel Core i7 processor, 16 GB RAM).
In the grid-space search, it takes the URCs model $\sim$16 milliseconds on average to complete a calculation. By comparison, a P2D simulation implemented in commercial software COMSOL requires an average of 432 seconds to complete. We also pit URCs against a state-of-the-art fast P2D solver PyBaMM\autocite{sulzer_python_2021}. When applied to the same grid search, PyBaMM consumes $\sim$10.2 seconds per simulation on average, which is more than 600 folds than the running time of the URCs model. The cathode/separator/anode stack is discretized by 100/20/115 grid points and the electrode particles by 20 points, which are typical for P2D simulations.

\subsubsection{Gradient-based Optimization}
\begin{figure}[H]
    \centering
    \includegraphics[width = 0.95\textwidth]{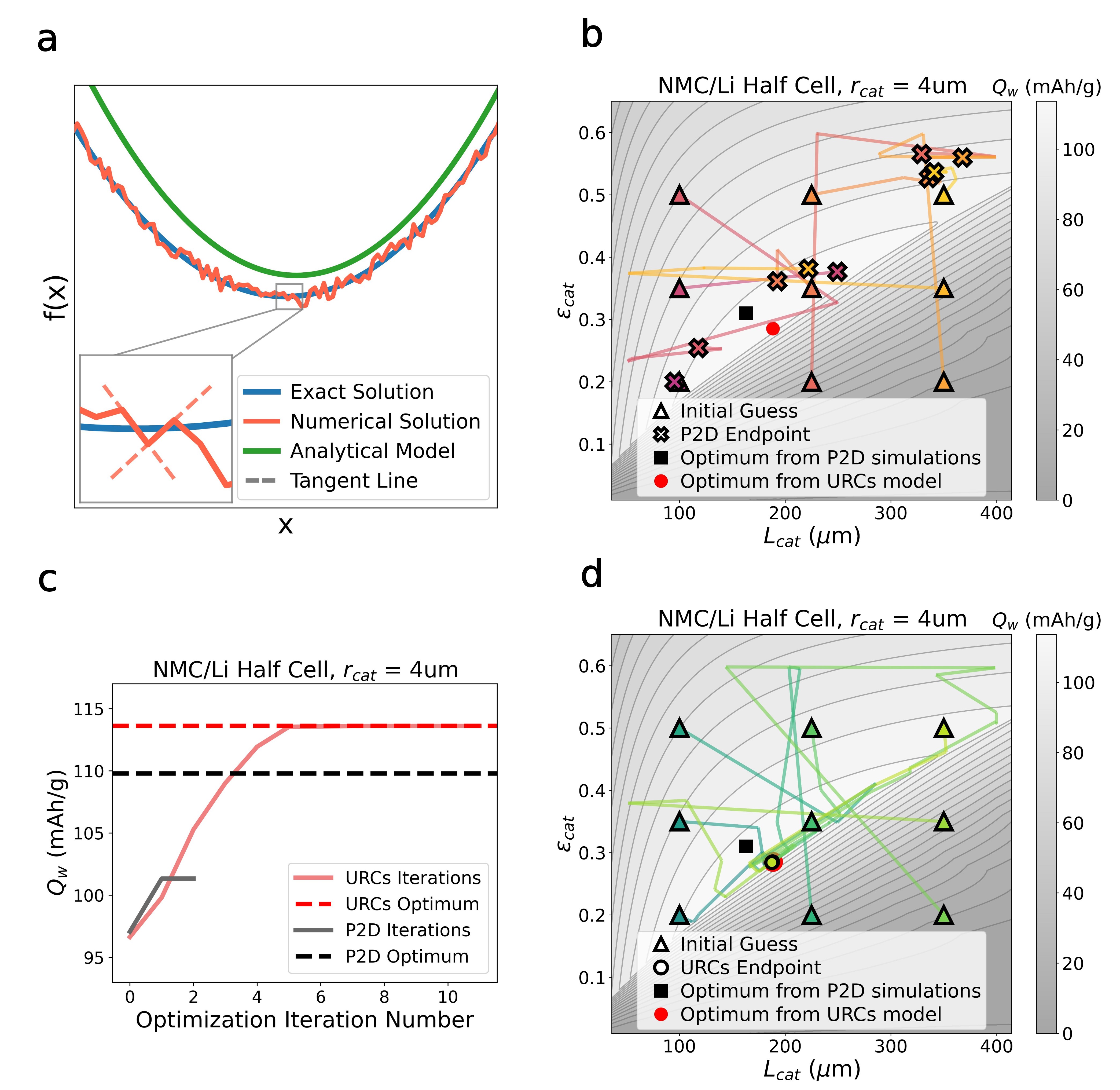}
        
    \caption{\textbf{Optimization of NMC half cells to maximize the cell level specific capacity $Q_w$ at 1C discharge with the URCs model and the gradient-based optimization method.} \\
    \textbf{a} Schematic of the different behaviors of the P2D simulation and the URCs model when used in gradient-based optimization. Numerical errors in P2D result in non-smooth objective function, leading to discontinuous first derivatives and failure to find the global optimum. The objective function evaluated by URCs is smoother, which enables the optimization to converge. \textbf{b} P2D-based optimization trajectories (solid lines) on the $L_{cat}$-$\epsilon_{cat}$ plane, superimposed on the contour plot of $Q_w$ generated by the URCs model. Searches from different initial guesses (triangle symbols) terminate at different endpoints (cross symbols) away from the global optimum departmentlocated by the P2D-based grid-space search (black square).    
    \textbf{c} Improvement of $Q_w$ during the optimization iteration process using P2D (black line) and URCs (red line). Initial guess is $L_{cat} = 100\mu$m and $\epsilon_{cat} = 0.35$. The black and the red dashed lines represent the global maximum  $Q_{w}$ from the grid-space search calculated by P2D and URCs, respectively.
    \textbf{d} URCs-based optimization trajectories (solid lines) show that searches from different initial guesses (triangle symbols) converge to the same global URCs optimum (red circle).
    }\label{fig:gradient}

\end{figure}

While we use the grid-space search to illustrate the computation efficiency of the URCs model in the last section, a perhaps even more notable advantage of the URCs model over P2D simulation lies in its compatibility with the gradient-based optimization methods, a family of the most widely used algorithms for finding the optimum of the objective function. In our test, we find that P2D simulation performs poorly with the gradient-based methods, which are usually unable to find the optimal cell parameters. In previous studies\autocite{dai_graded_2015}, derivative-free optimization algorithms were instead employed in conjunction with P2D despite the need for significantly more objective function evaluations than the gradient-based methods. P2D's unsatisfactory performance when used in optimization most likely stems from its inherent numerical errors, which generates inaccurate estimates of the first derivatives that cause the gradient-based search to fail\autocite{kolda_optimization_2003}. As illustrated in Figure \ref{fig:gradient}a, P2D simulation tends to produce non-smooth objective function due to the errors introduced by the discretization of the system and the differential algebraic equation solver. This causes its derivatives to be ill-behaved  and the search prone to divergence or being trapped in fictitious local minima. On the other hand, the analytical nature of the URCs model allows the objective function and its derivatives to be evaluated at much higher precision, making it excel in gradient-based optimization.

As a demonstration, we repeat the task of maximizing $Q_w$ at 1C discharge against $L_{cat}$ and $\epsilon_{cat}$ for half cells with $r_{cat}=4\mu$m by using the gradient-based methods implemented in MATLAB's \texttt{fmincon} function, which solves constrained optimization problems. In the optimization process, the upper and lower bounds for $L_{cat}$ and $\epsilon_{cat}$ are set to be [50$\mu$m, 400$\mu$m] and [0.15, 0.6], respectively, 
and we let the search start at nine different initial guesses from the different combinations of $L_{cat}\in\{100\mu\textrm{m}, 225\mu\textrm{m}, 350\mu\textrm{m}\}$ and $\epsilon_{cat}\in\{0.2, 0.35, 0.5\}$. Figure \ref{fig:gradient}b shows the initial guesses (triangular symbols), the successive steps (solid lines) and the final outcomes (cross symbols) of the search when P2D and the default optimization method (interior-point) and tolerance setting in \texttt{fmincon} are used. It can be seen that the optimization results from different initial guesses (triangular symbols) are all scattered in the parameter space and none of them is close to the global optimal configuration (black square) determined by the grid-space search. The optimization process is prone to premature termination as a result of being trapped in local minima. For example, Figure \ref{fig:gradient}c shows that the trial that begins at $(L_{cat}, \epsilon_{cat}) = (100\mu\textrm{m}, 0.35)$ ends in just three steps.

In contrast, the URCs model fares considerably better in the same task. As shown in \ref{fig:gradient}d, searches from all 9 starting points successfully converge to the true optimum identified by the grid-space search. Figure \ref{fig:gradient}c highlights the high efficiency of optimization with URCs. With the initial guess $(L_{cat}, \epsilon_{cat}) = (100\mu\textrm{m}, 0.35)$, the solver approaches the neighborhood of the optimum in just 5 iterations and converges after another 6 iterations. On average, the URCs model is evaluated for 85 times per optimization attempt (note that multiple function calls are made in each iteration). Using other gradient-based algorithms available in \texttt{fmincon} (e.g. sequential quadratic programming) yields similar performance. We note that the optimization outcomes show little sensitivity to the selected algorithm and the bounds of independent variables with discrepancy within 1$\text{\textperthousand}$. As a comparison, we also test the performance of combining the URCs model with a derivative-free method, the pattern search algorithm, which was used to optimize battery cell configuration with P2D simulations by Dai and Srinivasan\autocite{dai_graded_2015}. The algorithm is implemented in MATLAB's \texttt{patternsearch} function, and the default solver setting is used except raising the maximum iterations and function calls. An average of 4997 objective function evaluations are needed by the algorithm to find the optimum, which is 59 times of what is required by a gradient-based method.

\subsubsection{Hybrid Optimization Scheme}
To take advantage of the speed of the URCs model and further improve the prediction accuracy, we propose a hybrid approach to optimizing battery cell configurations. The idea is straightforward. Gradient-based optimization is first applied with the URCs model to quickly bring the search near the global optimum. A parameter sweep using P2D simulations is then carried out in the local neighborhood to locate the optimal parameters more accurately. We use the URCs to effectively reduce the domain size for P2D-based grid-space search, which is time- and resource-consuming. Based on the relative difference between the URCs and P2D results revealed in our test, it is reasonable to set the search domain within c.a. $\pm$15\% of the URCs-predicted optimal parameter values. 

\begin{figure}[!tbh]
    \centerline{
    \includegraphics[width = 1.1\textwidth]{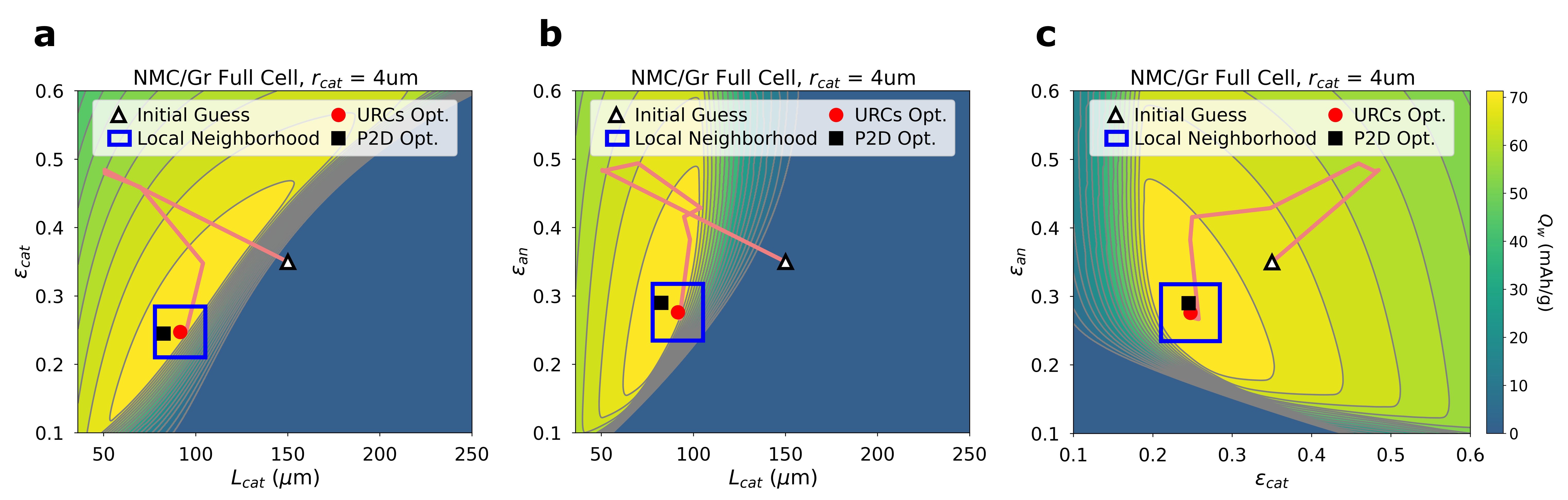}
    }
    \caption{\textbf{Optimization of NMC/Gr full cells to maximize the cell level specific capacity $Q_w$ at 1C discharge with hybrid optimization scheme.}\\
    The gradient-based optimization algorithm (trajectory indicated with the solid line) successfully converges to the optimum predicted by the URCs model (red circle), originating from an arbitrary initial guess (triangle symbol) of $L_{cat}=150\mu\textrm{m}$, $\epsilon_{cat}=0.35$, and $\epsilon_{an}=0.35$. A local grid-space search is performed around the URCs-estimated optimum (search region indicated with blue boxes) using P2D simulations. The local optimum predicted by P2D (black square) is subsequently verified to be the global optimum. \textbf{a}, \textbf{b}, and \textbf{c} illustrate the hybrid optimization scheme on the $L_{cat}$-$\epsilon_{cat}$, $L_{cat}$-$\epsilon_{an}$, and $\epsilon_{cat}$-$\epsilon_{an}$ planes, respectively. For each plot, the variable held constant maintains the same value as that of the P2D-generated global optimum.
    }\label{fig:3vars}
    
\end{figure}

To test the proposed hybrid scheme, we optimize the cathode thickness $L_{cat}$, cathode porosity $\epsilon_{cat}$ and anode porosity $\epsilon_{an}$ of an NMC/Gr full cell to maximize its specific capacity $Q_w$ at 1C discharge. The anode thickness $L_{an}$ is adjusted accordingly to give an anode:cathode capacity ratio of 1.1. Because of the increased number of design variables, performing a global grid-space search similar to Sec. \ref{sec:grid-space search} would require over 15000 P2D simulations (5415 coarse scan and 11466 refined scan), which are highly expensive computation-wise. Using the hybrid approach instead, we let the gradient-based search start at an arbitrary initial guess $(L_{cat}, \epsilon_{cat}, \epsilon_{an})=(150\mu\textrm{m}, 0.35, 0.35)$. It takes only 101 function evaluations and a few seconds of computation time for the optimizer to find the URCs optimum at $(L_{cat}^\textrm{opt*}, \epsilon_{cat}^\textrm{opt*}, \epsilon_{an}^\textrm{opt*})=(91.5\mu\textrm{m}, 0.247, 0.276)$ and $Q_w^{opt*}$ =  71.39 mAh/g. The optimization trajectory in the parameter space is visualized in Figure \ref{fig:3vars}a, b, and c, each illustrating a cross-sectional plane formed by two of the three independent parameters. The subsequent P2D scan is carried out in the neighborhood (blue boxes in Figure \ref{fig:3vars}) of $L_{cat}\in (77.8\mu\textrm{m}, 105.3\mu\textrm{m})$, $\epsilon_{cat}\in (0.210, 0.284)$ and $\epsilon_{an}\in (0.235, 0.318)$ with step size 2$\mu$m for $L_{cat}$ and 0.01 for $\epsilon_{cat}$ and $\epsilon_{an}$, resulting in a total of 728 simulations. The optimal configuration predicted by P2D is located at $(L_{cat}^\textrm{opt}, \epsilon_{cat}^{opt}, \epsilon_{an}^{opt})=(82.5\mu\textrm{m}, 0.245, 0.29)$ with $Q_w^{opt}$ = 69.79 mAh/g, which differ from the URCs optimum by 10.91\%, 0.98\%, 4.73\% and 2.3\%, respectively. The consistently good agreement between the URCs and P2D predictions even as the complexity of the optimization problem increases is the reason that we could limit the P2D-based search to a narrow region surrounding the URCs optimum in the parameter space to significantly reduce the computation cost without compromising the prediction accuracy.

\section{Conclusion}
In this study, we present a physics-based analytical model (the URCs model) for facile prediction of the battery cell performance under mixed kinetic control of electrolyte transport and solid-state diffusion. It is suitable for electrode materials such as NMC that remain as solid solution during lithium (de)intercalation. The URCs model simplifies the porous electrode theory by assuming pseudo-steady-state electrolyte transport and a uniform reaction distribution within the salt penetration zone. As a result, analytical expressions of the salt concentration and electrical potential distributions in the electrolyte could be derived, which are coupled to the analytical solution of the lithium solid diffusion equation in electrode particles to determine DoD as a function of the cell voltage. The URCs model exhibits very good agreement with the P2D simulations in predicting the discharge capacity, voltage curve and specific energy of battery cells though relatively large discrepancy is observed for NMC/Gr full cells at high rates because of the hybrid reaction behavior of the graphite anode. The power of the model in battery cell optimization is demonstrated. Its high computational speed enables the model to quickly scan the design variable space to reveal their effect on performance, which is very time-consuming for P2D simulations. While P2D cannot reliably find the global optimum when used in gradient-based optimization, the analytical nature of the URCs model makes it highly compatible with such optimization methods. We suggest that the URCs model could be combined with P2D simulations in a hybrid scheme to optimize the battery structure with both efficiency and accruacy. Overall, the light weight and versatility of the URCs model ready it for battery design tasks and real-time onboard applications.

\section*{Methods}
\subsection*{P2D Simulation}
The P2D simulations based on the porous electrode theory have been extensively discussed in various literature sources\autocite{doyle_modeling_1993, fuller_simulation_1994, ferguson_nonequilibrium_2012, thomas_mathematical_2002}. In this context, we offer a summary of the equations used in the P2D simulations, several of which have been previously presented in the derivation of the URCs model. Mass balance and current continuity in the electrolyte are described by Equations \ref{eq:masscon} and \ref{eq:currentcon}, respectively. Current continuity in the solid phase is given by
\begin{equation}
    \nabla\cdot\vec{i_s}(x)= F a_p(x) j_{in}(x) \label{eq:currentcon_solid}
\end{equation}
The intercalating flux at solid particle surface $j_{in}$ is governed by the Butler-Volmer equation:
\begin{equation}
    j_{in} = \frac{i_0}{F} \left[ \exp\left(-\frac{\alpha F\eta}{RT}\right) - \exp\left(\frac{\alpha F\eta}{RT}\right) \right] 
    \label{eq:BVeqn}
\end{equation} 
where $\alpha$ is the charge transfer coefficient. The expressions for the surface overpotential $\eta$ and the exchange current density $i_0$ are provided in Equations \ref{eq:etaEqn} and \ref{eq:j0}, respectively. Within the solid phase, current density and electrical potential are correlated by electrical conductivity $\sigma$:
\begin{equation}
    \vec{i_s}(x) = -\sigma \nabla \Phi_s \label{eq:ohmSolid}
\end{equation}
The relationship between the effective current density and potential in the electrolyte is expressed by Equation \ref{eq:ionicI}. Solid-state diffusion of Li in the active material is modeled as radial diffusion in spherical particles as described by Equation \ref{eq:sphDiffusion}. The P2D simulations that generated the figure data were implemented in COMSOL Multiphysics version 5.6. PyBaMM was also used for benchmarking purposes\autocite{sulzer_python_2021}. Values for the cell parameters used in the simulations are listed in Table \ref{Stab:P2Dparams}.

\subsection*{Data availability}
The MATLAB code that implements the URCs model is available at https://github.com/mingtang01/URCsBatteryModel

\newgeometry{top = 1.1in, bottom = 1.1in, left = 0.6in, right = 0.6in} 
\twocolumn
\begin{spacing}{1.5}
\tablefirsthead{\multicolumn{2}{c}{\textbf{List of Symbols Used}}\\}

\begin{supertabular}{p{0.22\linewidth} p{0.7\linewidth}} 

$a_{cat}/a_{an}$ & Volumetric surface area of cathode/anode [m\textsuperscript{-1}] \\ 
C & C rate \\
$C_\textrm{crit}$ & Critical C rate \\
$c_l$ & Salt concentration in electrolyte [mol·m\textsuperscript{-3}] \\
$c_{l0}$ & Initial salt concentration in electrolyte [mol·m\textsuperscript{-3}]\\
$c_s$ & Li concentration in electrode particles [mol·m\textsuperscript{-3}] \\
$\overline{c}_s$& Average Li concentration in electrode particles [mol·m\textsuperscript{-3}]\\
$c_{s0\,(cat/an)}$ & Initial Li concentration in electrode (cathode/anode) particles [mol·m\textsuperscript{-3}]\\
$c_{smax\, (cat/an)}$ & Maximum Li concentration in electrode (cathode/anode) particles [mol·m\textsuperscript{-3}]\\
$c_{ss\, (cat/an)}$ & Li concentration on electrode (cathode/anode) particle surface [mol·m\textsuperscript{-3}]\\
$\Delta c_{s\,(cat/an)}$ & Amount of Li (de)intercalated into an electrode (cathode/anode) particle\\
$D_{amb}$ & Ambipolar diffusivity of electrolyte [m\textsuperscript{2}·s\textsuperscript{-1}]\\
$D_{s\, (cat/an)}$ & Li diffusivity in active material [m\textsuperscript{2}·s\textsuperscript{-1}]\\
$\textrm{DoD}_{\,(cat/an)}$  & Depth of discharge (cathode/anode)\\
DoD\textsubscript{f} & Final depth of discharge\\
$E_A$ & Areal discharge energy [Wh/m\textsuperscript{2}]\\
$E_w$ & Cell-level specific energy [Wh/kg]\\
$F$ & Faraday constant (96485 C·mol\textsuperscript{-1})\\
$I$ & Applied current density [A·m\textsubscript{-2}]\\
$\vec{i}$ & Current density in liquid phase [A·m\textsuperscript{-2}]\\
$\vec{i_s}$ & Current density in solid phase [A·m\textsuperscript{-2}]\\
$i_0$ & Exchange current density of active material [A·m\textsuperscript{-2}]\\
$i_0^{Li}$ & Exchange current density on Li anode [A·m\textsuperscript{-2}]\\
$j_{in}$ & Reaction flux on active material surface [mol·m\textsuperscript{-2}·s\textsuperscript{-1}]\\
$k_0$ & Reaction rate constant [mol·m\textsuperscript{-2}·s\textsuperscript{-1}·(mol·m\textsuperscript{-3})\textsuperscript{}{-1.5}]\\
$L_{cat}$/$L_{sep}$/$L_{an}$ & Cathode/separator/anode thickness [m]\\
$L_\textrm{PZ}$ & Salt penetration zone thickness [m]\\
$Q_0$ & Areal capacity of the cathode [mAh/m\textsuperscript{2}]\\
$Q_w$ & Cell-level specific capacity [mAh/g]\\
$R$ & Gas constant (8.314 J·mol\textsuperscript{-1}·K\textsuperscript{-1})\\
$r$ & Spatial coordinate in electrode particle radial direction [m]\\
$r_{cat}$/$r_{an}$ & Cathode/anode particle radius [m]\\
SoC & State of charge \\
$T$ & Temperature [298 K]\\
$t$ & Time [s]\\
$t_+$ & Cation transference number in electrolyte \\
$U_{eq\,(cat/an)}$ & Equilibrium open-circuit potential of active material (cathode/anode) [V]\\
$X$/$x$ & Spatial coordinate in electrode thickness direction [m]\\
$1+\frac{\partial \ln{f_{\pm}}}{\partial \ln{c_l}}$ & Thermodynamic factor\\
$\alpha$ & Charge transfer coefficient \\
$\epsilon_{cat}$/$\epsilon_{sep}$/$\epsilon_{an}$ & Cathode/separator/anode porosity \\
$\eta_{\,(cat/an/Li)}$ & (Cathode/anode/Li metal) overpotential [V]\\
$\kappa$ & Ionic conductivity [S·m\textsuperscript{-1}]\\
$\nu_{cat}$/$\nu_{an}$ & Volumetric fraction of active material in cathode/anode\\
$\sigma$ & Electrical conductivity [S·m\textsuperscript{-1}]\\
$\tau_{cat}$/$\tau_{sep}$/$\tau_{an}$ & Cathode/separator/anode tortuosity \\
$\Phi_l$/$\Phi_{s\, (cat/an)}$ & Electrolyte/solid phase (cathode/anode) potential [V]\\
\end{supertabular}
\end{spacing}
\restoregeometry
\onecolumn

\section*{Acknowledgments}
H.W. is supported by DOE under project number DE-EE0006250 and Shell International Exploration and Production Inc. F.W. and M.T. acknowledge support from DOE Office of Basic Energy Sciences under project number DE-SC0019111. Simulations were partially performed on computing clusters at the Texas Advanced Computing Center (TACC) at the University of Texas.

\subsection*{Author Contributions}
M.T. conceived and supervised the research. H.W., F.W. and M.T. performed theoretical analysis and numerical calculations, discussed the results and wrote the manuscript. 

\subsection*{Conflicts of Interest}
The authors declare no competing financial interest.

\printbibliography
\pagebreak


\section*{Supplementary Materials}
\begin{refsection}

\setcounter{table}{0}
\renewcommand{\thetable}{S\arabic{table}}
\setcounter{figure}{0}
\renewcommand{\thefigure}{S\arabic{figure}}%

\begin{table}[H]
\centering
\caption{Parameters used in P2D simulations (unless otherwise stated)}\label{Stab:P2Dparams}%
\begin{tabular}{c|c|c|c}
\hline
Parameter & Symbol & \multicolumn{2}{c}{Value}\\\hline
\multicolumn{4}{c}{Electrode Properties}\\\hline
&&NMC&Graphite\\\hline
Cathode/anode particle radius (\textmu m) & $r_{cat}/r_{an}$ & See Text &  1 \\\hline
 Cathode/anode porosity& $\epsilon_{cat}$/$\epsilon_{an}$& 0.25&0.38\\\hline
 Cathode/anode tortuosity& $\tau_{cat}$/$\tau_{an}$& 1.3·$\epsilon_{cat}^{-0.8}$\citeSupp{usseglio-viretta_resolving_2018}&2.8·$\epsilon_{an}^{-1}$\citeSupp{usseglio-viretta_resolving_2018}\\ \hline
 Separator thickness (\textmu m)& $L_{sep}$& \multicolumn{2}{c}{25}\\\hline
 Separator porosity& $\epsilon_{sep}$& \multicolumn{2}{c}{0.55}\\\hline
 Separator tortuosity& $\tau_{sep}$& \multicolumn{2}{c}{$\epsilon_{sep}^{-0.5}$}\\ \hline
 Maximum Li concentration in active materials (mol·m\textsuperscript{-3})& $c_{smax}$& 49761&31507\\ \hline
 Initial Li concentration in active materials (mol·m\textsuperscript{-3})& $c_{s0}$& 22392&28986\\ \hline
 Li diffusivity in active materials (m\textsuperscript{2}·s\textsuperscript{-1})& $D_{s}$& 10\textsuperscript{-14}\citeSupp{amin_characterization_2016}&9·10\textsuperscript{-14}\citeSupp{srinivasan_design_2004}\\ \hline
 Electrode conductivity (S·m\textsuperscript{-1})& $\sigma_{s}$& 10\citeSupp{mao_multi-particle_2015}&100\citeSupp{srinivasan_design_2004}\\ \hline
 Reaction rate constant (mol·m\textsuperscript{-2}·s\textsuperscript{-1}·(mol·m\textsuperscript{-3})\textsuperscript{-1.5})& $k_0$& 3·10\textsuperscript{-11}\citeSupp{mao_multi-particle_2015}&3·10\textsuperscript{-11}\citeSupp{jokar_inverse_2016}\\ \hline
 Charge transfer coefficient& $\alpha$& \multicolumn{2}{c}{0.5}\\ \hline
 Exchange current density of Li anode (A·m\textsuperscript{-2})& $i_0^{Li}$& \multicolumn{2}{c}{20\citeSupp{mao_multi-particle_2015}}\\ \hline
 Equilibrium potential (V)&$U_{eq}$&\multicolumn{2}{c}{See note a}\\\hline
  \multicolumn{4}{c}{Electrolyte (1M LiPF\textsubscript{6} in PC:EC:DMC=10:27:63 volume ratio) properties}\\\hline
   Initial salt concentration (mol·m\textsuperscript{-3})&$c_{l0}$&\multicolumn{2}{c}{1000\citeSupp{valoen_transport_2005}}\\\hline
    Transference number of cations&$t_+$&\multicolumn{2}{c}{0.38\citeSupp{valoen_transport_2005}}\\\hline
     Thermodynamic factor&$1+\frac{\partial \ln{f_{\pm}}}{\partial \ln{c_l}}$&\multicolumn{2}{c}{1\citeSupp{valoen_transport_2005}}\\\hline
 Ambipolar diffusivity (m\textsuperscript{2}·s\textsuperscript{-1})& $D_{amb}(c_l)$& \multicolumn{2}{c}{See note b}\\\hline
 Ionic conductivity (S·m\textsuperscript{-1})& $\kappa(c_l)$& \multicolumn{2}{c}{See note c}\\ \hline
\end{tabular}

\end{table}
Note:
\begin{enumerate}[label=\textbf{\alph*.}]
\item The equilibrium potential profiles of NMC is extracted from Figure 2 in Ref.\cite{smekens_modified_2015}. The equilibrium potential of graphite is
adopted from Ref.\cite{srinivasan_design_2004} as
\begin{align*}
    &U_{eq, an}(x) = 0.124 + 1.5\exp{(-70x)}-0.0351\tanh{(\frac{x-0.286}{0.083})}-0.0045\tanh{(\frac{x-0.9}{0.119})}\\
    &-0.035\tanh{(\frac{x-0.99}{0.05})}-0.0147\tanh{(\frac{x-0.5}{0.034})}-0.102\tanh{(\frac{x-0.194}{0.142})}\\
    &-0.022\tanh{(\frac{x-0.98}{0.0164})}-0.011\tanh{(\frac{x-0.124}{0.0226})}+0.0155\tanh{(\frac{x-0.105}{0.029})}
\end{align*}
\item Calculated with tabulated parameter values in Ref.\cite{valoen_transport_2005} by:
\begin{equation*}
     D_{amb}^*(c_{l}^*) = 10^{\displaystyle -4.43-54/(T-(229+5c_l^*))-0.22c_l^*}
\end{equation*}
$D_{amb}^*$ is the ambipolar diffusivity measured in [cm\textsuperscript{2}/s], and $c_l^*$ is the salt concentration in electrolyte measured in [mol/L].
\item Calculated with tabulated parameter values in Ref.\cite{valoen_transport_2005} by:
\begin{equation*}
    \kappa^*(c_{l}^*) = c_{l}^* (-10.5+0.668 c_{l}^* +0.494 {c_{l}^*}^2+0.0740 T-0.0178 c_{l}^* T-8.86\textrm{e}-4{ c_{l}^* }^2 T-6.96\textrm{e}-5 T^2+2.8\textrm{e}-5 c_{l}^*  T^2)^2
\end{equation*}
$\kappa^*$ is the ionic conductivity measured in [mS/cm], and $c_l^*$ is the salt concentration in electrolyte measured in [mol/L].
\end{enumerate}

\pagebreak
\begin{table}[H]
\centering
\begin{tabular}{c|c|c|c|c}
\hline
Parameter & Symbol & \multicolumn{3}{c}{Value}\\\hline
 & & NMC& Lithium&Graphite\\\hline
Density of cathode active material (g·cm\textsuperscript{-3})& $\rho_{cat}$& 4.77& -&-\\\hline
 Density of anode active material (g·cm\textsuperscript{-3})& $\rho_{an}$& -& 0.534&2.27\\\hline
 Volume capacity of cathode active material (mAh·cm\textsuperscript{-3})& $Q_V^0$& 734& -&-\\\hline
 Anode/cathode capacity ratio& $R_{an/cat}$& -& 1.25&1.1\\\hline
 Density of separator (g·cm\textsuperscript{-3})& $\rho_{sep}$& \multicolumn{3}{c}{0.946}\\\hline
  Density of electrolyte (g·cm\textsuperscript{-3})& $\rho_{electrolyte}$& \multicolumn{3}{c}{1.3}\\\hline
   Density of copper (g·cm\textsuperscript{-3})& $\rho_{Cu}$& \multicolumn{3}{c}{8.96}\\\hline
    Density of aluminum (g·cm\textsuperscript{-3})& $\rho_{Al}$& \multicolumn{3}{c}{2.7}\\\hline
    Current collector thickness (\textmu m) & $L_{cc}$ & \multicolumn{3}{c}{15 (double-side coating assumed)}\\\hline
\end{tabular}
\caption{Cell component properties used in calculating the cell-level specific capacity $Q_{w}$ and energy $E_{w}$}\label{Stab:QwParams}%

\end{table}

\pagebreak
\printbibliography[heading=subbibintoc]
\end{refsection}
\end{document}